\documentclass[11pt]{article}
\usepackage{hyperref}
\usepackage[english]{babel}
\usepackage{microtype}
\usepackage{graphicx} 
\usepackage{epstopdf} 
\usepackage{amssymb}
\usepackage{amsmath}
\usepackage{fixme}
\usepackage{epsf} 
\usepackage{rotating} 
\usepackage{pstricks}
\usepackage{color} 
\usepackage{subfigure} 
\usepackage[small]{caption}
\usepackage{multirow} 
\usepackage{color}
\usepackage[top=2cm,left=2cm,right=2cm]{geometry}
\usepackage[numbers,square]{natbib} 
\usepackage{color}

\def\slashed{\ds}

\def\g{\gamma}

\def\ds#1{#1\kern-1ex\hbox{/}} 
\def\dsh{h\kern-1.2ex /}

\newcommand{\bea}{\begin{eqnarray}} 
\newcommand{\eea}{\end{eqnarray}}

\def\beq{\begin{equation}} 
\def\eeq{\end{equation}}

\def\beqn{\begin{eqnarray}}
\def\eeqn{\end{eqnarray}} 
\def\ba{\begin{eqnarray}}
\def\ea{\end{eqnarray}}

 \newcommand{\beqa}{\begin{eqnarray}}
\newcommand{\eeqa}{\end{eqnarray}}

\begin{document}

\begin{center} 
\vspace{4.cm}

{\bf \large Relic Densities of Dark Matter in the U(1)-Extended NMSSM\\
and the Gauged Axion Supermultiplet}

\vspace{1cm}

{\bf$^a$Claudio Corian\`{o},\,$^b$Marco Guzzi and $^a$Antonio Mariano}

\vspace{1cm}

{\it$^a$ Dipartimento di Matematica e Fisica\\ Universit\`{a} del Salento \\ and INFN
Sezione di Lecce, Via Arnesano 73100 Lecce,
Italy\footnote{claudio.coriano@unisalento.it,
antonio.mariano@unisalento.it} }\\ \vspace{.5cm} {\it$^b$Department of
Physics, Southern Methodist University, \\Dallas TX 75275,
USA\footnote{mguzzi@physics.smu.edu}\\} \vspace{.5cm} 

\begin{abstract}

We compute the dark matter relic densities of neutralinos and axions in a supersymmetric model with a gauged anomalous $U(1)$ symmetry. The model is a variant of the USSM (the
$U(1)$ extended NMSSM), containing an extra $U(1)$ symmetry and an
extra singlet in the superpotential respect to the MSSM, where gauge invariance 
is restored by Peccei-Quinn interactions using a St\"uckelberg multiplet. This approach
introduces an axion ($\textrm{Im}\, b$) and a saxion ($\textrm{Re}\,
b$) in the spectrum and generates an axino component for the
neutralino. The St\"uckelberg axion ($\textrm{Im}\, b$) develops a
physical component (the gauged axion) after electroweak symmetry
breaking. We classify all the interactions of the
Lagrangian and perform a complete simulation study of the
spectrum, determining the neutralino relic densities using micrOMEGAs. 
We discuss the phenomenological implications
of the model analyzing mass values for the axion
from the milli-eV to the MeV region. These depend sensitively on the value of $\tan\beta$. The possible scenarios that we analyze are significantly constrained by a combination of WMAP data,
the exclusion limits from direct axion searches and the veto on late
entropy release at the time of nucleosynthesis.

\end{abstract}
\end{center} \newpage

\section{Introduction} Axions have been studied along the years both
as a realistic attempt to solve the strong CP problem
\cite{Peccei:1977ur,Peccei:2006as},\cite{Weinberg:1977ma,Wilczek:1977pj,Dine:1981rt,Zhitnitsky:1980tq,Kim:1979if}\cite{Shifman:1979if},
to which they are closely related, but also as a possible candidate to
answer more recent puzzles in cosmology, such as the origin of dark
energy, whose presence has found confirmation in the study of Type I
supernovae \cite{Riess:1998cb,Perlmutter:1998np}. In this second case
it has been pointed out that they can contribute to the vacuum energy,
a possibility that remains realistic if their mass $m_a$ - which in
this case should be $\sim 10^{-33 }$ eV and smaller - is of
electroweak \cite{Nomura:2000yk} and not of QCD origin. In this case
they differ significantly from the standard (Peccei-Quinn, PQ)
invisible axion.

According to this scenario, the vacuum misalignment (see
\cite{Sikivie:2006ni, Chang:1998ys} for a discussion in the PQ case)
induced at the electroweak scale would guarantee that the degree of
freedom associated with the axion field remains frozen, rolling down
very slowly towards the minimum of the non-perturbative instanton
potential, with $m_a$ much smaller than the current Hubble rate.
Given the rather tight experimental constraints which have
significantly affected the parameter space (axion mass and gauge
couplings) for PQ axions
\cite{Arik:2008mq,Asztalos:2009yp,Duffy:2009ig}, the study of these
types of fields has also taken into account the possibility to evade
the current bounds \cite{Raffelt:2006cw, Visinelli:2009zm}. These are
summarized both into an upper and a lower bound on the size of $f_a$,
the axion decay constant, which sets the scale of the misalignment
angle $\theta$, defined as the ratio of the axion field ($a$) over the
PQ scale $v_{PQ}$ ($v_{PQ}\sim f_a$).

Axion-like particles can be reasonably described by pseudoscalar
fields characterized by an enlarged parameter space for mass and
couplings, with a direct coupling to the gauge fields (of the form $a
F\tilde{F}$) whose strength remains unrelated to their mass. They have
been at the center of several recent and less recent studies (see for
instance \cite{Ahlers:2007qf,Ahlers:2007rd}
\cite{DeAngelis:2007yu,DeAngelis:2007dy,Mirizzi:2009aj,Visinelli:2009zm,Berezhiani:1992rk,Berezhiani:1991uj}). They
are supposed to inherit most of the properties of a typical invisible
axion - a PQ axion - while acquiring some others which are not allowed
to it.
 
We recall that the axion mass (which in the PQ case is
$O(\Lambda_{QCD}^2/f_a)$ and the axion coupling to the gauge fields
are indeed related by the same constant $f_a$.  In the PQ case $f_a$
($\sim 10^{10}-10^{12}$ GeV) makes the axion rather light ($\sim
10^{-3}-10^{-5}$ eV) and also very weakly coupled. The same (large)
scale plays a significant role in establishing the axion as a possible
dark matter candidate, contributing significantly to the relic
densities of cold dark matter.  A much smaller value of $f_a$, for
instance, would diminish significantly the axion contribution to cold
dark matter due to the suppression of its abundance ($Y_\chi$) which
depends quadratically on $f_a$.

It is quite immediate to realize that the gauging of the axionic
symmetries by introducing a local anomalous $U(1)$ - inherited from
an underlying anomalous structure, i.e.\ a gauge anomaly - allows to
leave the mass and the coupling of the axion to the gauge fields
unrelated \cite{Coriano:2007fw,Coriano:2007xg}, offering a natural
theoretical justification for the origin of axion-like particles. We
just recall that effective low energy models incorporating gauged PQ
interactions emerge in several string and supergravity constructions,
for instance in orientifold vacua of string theory and in gauged
supergravities (see for instance
\cite{DeRydt:2007vg,Derendinger:2007xp}).

The analysis that we perform in this work has the goal to capture the
relevant phenomenological features of the axions present in this class
of models, extending a previous study presented in a
non-supersymmetric context \cite{Coriano:2010py}. We will be
following, as in previous studies, a bottom-up approach. This allows
to identify the low energy effective action on the basis of a rather
simple operatorial structure typical of anomalous abelian models. The
theory is then fixed by the condition of gauge invariance of the
anomalous effective action, amended by operators of dimension-5
(Wess-Zumino or PQ-like terms) which appear in the action suppressed
by a suitable scale, the St\"uckelberg mass ($M_{St}$).

The introduction of the St\"uckelberg multiplet (or axion multiplet),
while necessary for the restoration of gauge invariance, is in general
expected to raise some concerns at cosmological level because of the
presence, among its components, of a scalar modulus, the saxion. In
supersymmetric (ordinary) PQ formulations this has a mass of the order
of the weak scale or smaller and poses severe problems to the standard
cosmological scenario. A late time decay of this particle, for
instance, could cause an entropy release with a low reheating
temperature ($T_{RH} < 5 \,\textrm{MeV}$) which is unacceptable for
nucleosynthesis.  Just for comparison, we mention that in the case of
string moduli, for instance, the interaction of these states with the
rest of the fields of the low energy spectrum is suppressed by the
Planck scale. In turn, this forces the mass of these states to be
quite large (100 TeV or so) \cite{deCarlos:1993jw, Banks:2002sd} in
order to enhance the phase space for their decay, for a similar
reason.

In our construction the scalar modulus of the axion multiplet acquires
a mass of the order of the St\"uckelberg scale and has sizeable
interactions with the other fields of the model, thereby decaying
pretty fast ($\sim 10^{-23} s$). Therefore, smaller values of its mass
- in the TeV range - turn out to be compatible with the standard
scenario for nucleosynthesis.

\subsection{The USSM-A } 
Non-supersymmetric versions of the class of
models that we are going to analyze have been discussed in details in
\cite{Coriano':2005js,Coriano:2007fw,Coriano:2007xg}. Recently
\cite{Coriano:2008xa,Coriano:2008aw}, an extension of a specific
supersymmetric model, the USSM (the $U(1)$-extended Next-to-Minimal
Supersymmetric Standard Model of \cite{Cvetic:1997ky}) has been
presented, in which the $U(1)$ symmetry is anomalous. This model
supports an axion-like particle in its spectrum. It has been named the
``USSM-A'', to recall both its supersymmetric origin and its anomalous
abelian gauge structure. It is also close to a similar $U(1)^\prime$
extension of the MSSM (the $U(1)^\prime$ Minimal Supersymmetric
Standard Model) \cite{Anastasopoulos:2008jt}, which supports an axino
component among the interaction eigenstates of the neutralino sector,
but not a gauged axion, due to the structure of the MSSM
superpotential. The study of relic densities in this model have been
performed in \cite{Fucito:2008ai,Lionetto:2009dp}. In the
non-supersymmetric case the identification of a physical axion in the
spectra of these models has been discussed in detail in
\cite{Coriano':2005js}, a realization called ``the Minimal Low Scale
Orientifold Model'' or MLSOM for short.

Both in the USSM-A and in the model of \cite{Anastasopoulos:2008jt},
the extra $U(1)$ symmetry takes an anomalous form and the violation of
gauge invariance requires supersymmetric PQ interactions, with a
St\"uckelberg supermultiplet for the restoration of the gauge
symmetry. The extra gauge boson of the anomalous $U(1)$ symmetry is
massive and in the St\"uckelberg phase, as in previous
non-supersymmetric constructions
\cite{Coriano':2005js,Coriano:2007fw,Coriano:2007xg}. As shown in the
case of the MLSOM, axion-like particles appear in the CP-odd spectrum
of these theories whenever Higgs-axion mixing \cite{Coriano':2005js}
occurs. For this reason in this work we will be using the term
``gauged supersymmetric axion'' (or axi-Higgs, denoted equivalently as
$\chi$ or $H_0^5$) to refer to this state.

As we have mentioned in the introduction, we will follow a minimal
approach. This approach allows to define an effective theory on the
basis of 1) an assigned gauge structure (the number of anomalous
abelian interactions); 2) some conditions of anomaly cancellation and
gauge invariance of the effective Lagrangian; 3) the choice of a
suitable value of the St\"uckelberg mass scale characterizing the
range in which the description of these effective models is compatible
with unitarity \cite{Coriano:2008pg}. As in a previous analysis for
the LHC in the MLSOM \cite{Coriano:2009zh}, we will first stress on the
general features of these models, deriving the defining conditions for the counterterms which 
appear in the structure of the effective action, before moving to a specific realization with a selected 
charge assignment. In our simulations we have found that the dependence of the results on the choices of the independent charges is, however, extremely mild. In this respect the properties that we are able to 
extrapolate from this class of models - even with a single charge assignment - are pretty general and depend quite sensitively only on the choice of the St\"uckelberg 
mass $M_{St}$ and the MSSM Higgs vev ratio $\tan \beta$.

In this section we will focus on the axion/saxion Lagrangian, leaving a 
general discussion of the various contributions to an appendix. It is given by
\begin{align} {\cal L}_{axion/saxion}={\cal L}_{St} +{\cal L}_{WZ}
\end{align} where ${\cal L}_{St}$ is the supersymmetric version of the
St\"uckelberg mass term \cite{Kors:2004ri}, while ${\cal L}_{WZ}$
denotes the WZ counterterms responsible for the axion-like nature of
the pseudoscalar $b$. Specifically
\begin{align} 
{\cal L}_{St}&=\frac{1}{2}\int d^{4}\theta (\hat{{\bf
b}} +\hat{{\bf b}}^{\dagger}+\sqrt{2} M_{St} \hat{B})^{2}\nonumber\\ {\cal
L}_{WZ} &= -\frac{1}{2}\int d^{4}\theta\left\lbrace
\left[ \frac{c_{G}}{M_{St}} \,\textrm{Tr}({\cal G} {\cal G})\hat{{\bf b}}
+ \frac{c_{W}}{M_{St}} \,\textrm{Tr}(W W)\hat{{\bf b}}
\right.\right.  \nonumber\\
&\left.\left.+\frac{c_{Y}}{M_{St}}\hat{{\bf b}}W^{Y}_{\alpha}
W^{Y,\alpha} +\frac{c_{B}}{M_{St}}\hat{{\bf
b}}W^{B}_{\alpha}W^{B,\alpha}+\frac{c_{YB}}{M_{St}}\hat{{\bf
b}}W^{Y}_{\alpha}W^{B,\alpha}\right] \delta(\bar{\theta}^{2})
+h.c.\right\rbrace,
\end{align} 
where we have denoted with $\cal{G}$ the supersymmetric
field-strength of $SU(3)_c$, with $W$ the supersymmetric
field-strength of $SU(2)$, with $W^{Y}$ and with $W^{B}$ the
supersymmetric field-strength of $U(1)_{Y}$ and $U(1)_{B}$
respectively. The Lagrangian ${\mathcal L}_{St}$ is invariant under
the $U(1)_B$ gauge transformations
\begin{align} \delta_B
\hat{B}=&\hat{\Lambda}+\hat{\Lambda}^\dagger\nonumber\\ \delta_B
\hat{\bf{b}}=& - 2 M_{St} \hat{\Lambda}
\end{align} where $\hat{\Lambda}$ is an arbitrary chiral
superfield. So the scalar component of $\hat{\bf{b}}$, that consists
of the saxion and the axion field, shifts under a $U(1)_B$ gauge
transformation.\\ The coefficients $c_I\equiv(c_G,c_W, c_Y, c_B,
c_{YB})$ are dimensionless, fixed by the conditions of gauge
invariance, and are functions of the free charges $B_i$ of the model
(as shown below in Eq.~(\ref{anomalies1})). Extracting the group factors we have
\begin{align} 
&c_{B} = - \frac{{\cal A}_{BBB}}{384 \pi^2} \qquad
\qquad c_{Y} = - \frac{{\cal A}_{BYY}}{128 \pi^2 }\qquad \qquad c_{YB}
= - \frac{{\cal A}_{BBY}}{128 \pi^2 } \nonumber \\ &c_{W} = -
\frac{{\cal A}_{BWW}}{64 \pi^2} \qquad \qquad c_{G} = - \frac{{\cal
A}_{BGG}}{64 \pi^2}.
\label{coeffc}
\end{align} 
The coefficients ${\cal A}$ are defined by the
conditions of gauge invariance of the effective action, related to the
anomalies $\left\{U(1)_{B}^3\right\}$,
$\left\{U(1)_{B},U(1)_{Y}^2\right\}$,
$\left\{U(1)_{B}^2,U(1)_{Y}\right\}$,
$\left\{U(1)_{B},SU(2)^2\right\}$,
$\left\{U(1)_{B},SU(3)^2\right\}$. 
Using
the conditions of gauge invariance these coefficients assume the form
\begin{align} 
{\cal A}_{BBB}&= -3B_{H_1}^3 - 3 B_{H_1}^2(3 B_L + 18
B_Q - 7 B_S) - 3 B_{H_1} (3 B_L^2 +(18 B_Q - 7 B_S) B_S ) \nonumber\\
& + 3 B_L^3 + B_S (27 B_Q^2 - 27 B_S B_Q + 8 B_S^2) \nonumber\\ {\cal
A}_{BYY}&= \frac{1}{2}(-3 B_L - 9 B_Q + 7 B_S) \nonumber\\ {\cal
A}_{BBY}&= 2 B_{H_1} (3 B_L + 9 B_Q - 5 B_S) + (12 B_Q - 5 B_S) B_S
\nonumber\\ {\cal A}_{BWW}&= \frac{1}{2}(3 B_L + 9 B_Q - B_S)
\nonumber\\ {\cal A}_{BGG}&= \frac{3}{2} B_S.
\label{anomalies1}
\end{align} 
We have expressed all the anomaly equations in terms of 4
charges, $B_{i}\equiv (B_{H_1}, B_S, B_Q, B_L)$ ordered from 1 to 4
(left to right), which are defined in Tab. \ref{fieldcontent}. Notice that these charges can be taken as fundamental
parameters of the model. Their independent variation allows to scan
the entire spectra of these models with no reference to any specific
construction. These relations appear in the anomalous variation
$(\delta_B)$ of the supersymmetric 1-loop effective action of the
model, which forces the introduction of supersymmetric PQ-like
interactions (WZ terms) for its overall vanishing.  Formally we have
the relation
\begin{equation} 
\delta_B(B_i){\mathcal S}_{\it 1 loop} + \delta_{B}(
c_I(B_i)){\mathcal S}_{WZ} =0,
\label{formalEQ}
\end{equation} 
where the anomalous variation can be parameterized by
the 4 charges $B_i$ together with the coefficients $c_J( B_i)$ in
front of the WZ counter-terms.  In these notations, the uppercase
index $J$ runs over all the 5 mixed-anomaly conditions $B^3, B Y^2 ,
B^2 Y, B W^2, B G^2 $, ordered from 1 to 5 (left to right).\\
Before coming to the definition of the charge assignments we pause for a remark. 
As we are going to show in the next sections, the scalar potential takes a nonlocal form 
unless all the anomaly coefficients in Eq.~\ref{anomalies1} are zero. Such potential can however be expanded in powers of $\textrm{Re}b/M_{St}$, and as such these contributions turn out to be irrelevant if $M_{St}$ is a very large scale. The situation is rather different if $M_{St}$ is bound to lay around the 1 TeV region, where the potential could actually develope a singularity. In fact, in this case, it is in general expected that a singular potential will soon dominate the dynamics of the model.

We will give the explicit expression of the $D$-terms for a general choice of the counterterms.  
The function ($f$) which allows to identify all the
charges in terms of the free ones is formally given by 
\bea f(B_Q, B_L, B_{H_1}, B_S)= ( B_Q, B_{U_R}, B_{D_R}, B_L, B_R, B_{H_1},
B_{H_2}, B_S).
\label{charges1} \eea 
These depend only upon the 4 free parameters $B_Q, B_L$, $B_{H_1}$ and $B_S$. In our analysis, the charges of
Eq.~(\ref{charges1}) have been assigned as 
\begin{align} 
f(2, 1, -1, 3)= 
(2, 0, -1, 1, 0, -1, -2, 3).  
\label{fixcharge}
\end{align}

As we have already mentioned, the dependence of our results on this
choice of thi parametric charges is very small. Instead, as we will
see, the relevant parameters of our analysis turn out to be: 1) the
anomalous coupling of the gauge boson $g_B$, which controls the decay
rate of the saxion and of the axion and 2) the St\"uckelberg mass. 

\section{Axions, saxions and all orders interactions}
The contributions of the axion and saxion to the total Lagrangian
are derived from the combination of the St\"uckelberg and Wess-Zumino
terms. The complete axion/saxion Lagrangian expressed in terms of
component fields is given by 
\begin{align} 
{\cal L}_{axion/saxion}\equiv{\cal
L}_{St} + {\cal L}_{WZ} 
\end{align}
and contains a mixing among the $D$-terms
which is rather peculiar, as we are going to show. The off-shell
expression of this Lagrangian is given by
\begin{align} 
&{\cal L}_{axion/saxion}=\frac{1}{2}\left( \partial_{\mu}\textrm{Im}\,b +
M_{st}B_{\mu}\right)^{2}+\frac{1}{2}\partial_\mu\textrm{Re}b\,\partial^\mu\textrm{Re}b
+\frac{i}{2}\psi_{\bf b}\sigma^{\mu}\partial_{\mu}\bar{\psi_{\bf b}}
+\frac{i}{2}\bar{\psi_{\bf
b}}\bar{\sigma}^{\mu}\partial_{\mu}\psi_{\bf b} +F_{\bf b}^{\dagger}F_{\bf
b}+ {\cal L}_{axion, i}
\end{align}
where the expression of ${\cal L}_{axion, i}$ is quite lengthy and can be found in the appendix (Eq. (\ref{laxion})).

The equation of motion for the auxiliary field $F_{\bf b}$ can be derived quite immediately and give for the $F-$term of the St\"uckelberg field the expression
\begin{align} 
F_{\bf b}=
-\frac{1}{16}\frac{c_{G}}{M_{St}}\bar{\lambda}^a_{G}\bar{\lambda}^a_{G}
-\frac{1}{16}\frac{c_{W}}{M_{St}}\bar{\lambda}^i_{W}\bar{\lambda}^i_{W}
-\frac{1}{2}\frac{c_{Y}}{M_{St}}\bar{\lambda}_{Y}\bar{\lambda}_{Y}
-\frac{1}{2}\frac{c_{B}}{M_{St}}\bar{\lambda}_{B}\bar{\lambda}_{B}
+\frac{1}{2}\frac{c_{YB}}{M_{St}}\bar{\lambda}_{Y}\bar{\lambda}_{B}.
\end{align}
One important feature of the supersymmetric model is that $b$ is a
complex field with its real and imaginary parts. While $\textrm{Im}\,
b$ may appear in the CP-odd part of the scalar sector and undergoes
mixing with the Higgs sector, its real part, $\textrm{Re}\, b$, the
saxion (or scalar axion) before the EW symmetry breaking, has a mass exactly 
equal to the St\"uckelberg mass, as expected from supersymmetry. 
We just recall that in the
absence of SUSY breaking parameters, the components of the
St\"uckelberg multiplet form, together with the vector multiplet of
the anomalous gauge boson, a massive vector multiplet of mass
$M_{St}$. This is composed of the massive anomalous gauge boson, whose
mass is given by the St\"uckelberg term, the massive saxion and a
massive Dirac fermion of mass $M_{St}$.  The fermion is obtained by
diagonalizing the 2-dimensional mass matrix constructed in the basis
of $\lambda_B$ - the gaugino from the vector multiplet $\hat{B}$ - and
$\psi_b$, which is the axino of the St\"uckelberg multiplet. The
diagonalization of this matrix trivially gives two Weyl eigenstates of
the same mass $M_{St}$, which can be assembled into a single massive
Dirac fermion of the same mass. Notice that in this re-identification
of the degrees of freedom contained in $\hat{\bf b}$ and in the vector
multiplet $\hat{B}$, $\textrm{Im}\, b$ takes the role of a
Nambu-Goldstone mode and can be gauged away.

The saxion has typical interactions of the form $\textrm{Re}\, b\, F_i
F_j$, with the gauge fields which have mixed-anomalies with $U(1)_B$,
beside non-polynomial interactions with the remaining fields of the
theory. As we are going to elaborate, this features shows up because
of the presence of terms consisting of the product of two $D$ fields 
and the saxion.
To clarify this point, we recall that the general Lagrangian contains
a supersymmetric Wess-Zumino term of the form
\begin{equation} 
{\cal L}_{WZ,Y} = -\frac{1}{2}\int d^{4}\theta\;
\frac{c_{Y}}{M_{St}}\hat{{\bf b}}W^{Y}_{\;\alpha}W^{Y\;\alpha}\;
\delta(\bar{\theta}^{2}) +h.c.
\end{equation} 
which gives, after the expansion in components, a term
proportional to
\begin{equation} 
\frac{c_{Y}}{M_{St}}\;\textrm{Re}b\;D_Y D_Y.
\end{equation} 
This kind of terms, once that the equations of motion (EOM) of the auxiliary fields D are calculated
and substituted back into the lagrangian,
give the non-polynomial form of the potential. Furthermore, from the WZ term
corresponding to the anomaly $BBY$ (which is the term proportional to $c_{YB}$), we get 
a term proportional to $\textrm{Re}b\;D_Y D_B$ so that the EOM for the abelian D fields 
are coupled.\\ 
The derivation of such equations involves all the terms of the Lagrangian discussed in appendix~\ref{thelagrangian}.
We obtain
\begin{align} 
D_{B,OS}=&\frac{1}{12+12\sqrt{2}\textrm{Re}b(c_B + c_Y )/M_{St}  -
6 \textrm{Re}b^2(c_{YB}^2 - 4 c_Y c_B)/M_{St}^2}\notag\\
&\bigg\{ \left[ 2 \frac{c_B}{M_{St}} \left(\sqrt{2} + 2 \frac{c_Y}{M_{St}} 
\textrm{Re}b\right) - \frac{c_{YB}^2}{M_{St}^2} \textrm{Re}b\right] 
(-3 i \lambda_B\psi_{\bf b} + h.c.)-
12 M_{St} \textrm{Re}b 
- 12\sqrt{2}c_Y \textrm{Re}b^2 
+ \notag\\
& 12 g_B \left(1+\sqrt{2}\frac{c_Y}{M_{St}}\textrm{Re}b\right) 
\big( B_S S^\dagger S + B_{H_1} H_1^\dagger H_1 +  B_{H_2} H_2^\dagger H_2 + 
B_{D_R}\tilde{D}_R^\dagger\tilde{D}_R + B_{U_R}\tilde{U}_R^\dagger\tilde{U}_R + \notag\\
& B_{Q}\tilde{Q}^\dagger\tilde{Q} + B_{R}\tilde{R}^\dagger\tilde{R} + B_{L} \tilde{L}^\dagger \tilde{L}\big) + 3 \sqrt{2} \frac{c_{YB}}{M_{St}}
(i \lambda_Y\psi_{\bf b} + h.c.) + \notag\\
&\sqrt{2} \frac{c_{YB}}{M_{St}} g_Y \textrm{Re}b
\big( -3 H_1^\dagger H_1 + 3 B_{H_2} H_2^\dagger H_2
+ 2 \tilde{D}_R^\dagger\tilde{D}_R - 4 \tilde{U}_R^\dagger\tilde{U}_R +
\tilde{Q}^\dagger\tilde{Q} + 6 \tilde{R}^\dagger\tilde{R} - 
3 \tilde{L}^\dagger \tilde{L}\big)
\Big\}\notag\displaybreak[0]\\
D_{Y,OS}=&\frac{1}{12+12\sqrt{2}\textrm{Re}b(c_B + c_Y )/M_{St}  -
6 \textrm{Re}b^2(c_{YB}^2 - 4 c_Y c_B)/M_{St}^2}\notag\\
&\Big\{3\;\frac{c_{YB}^2}{M_{St}^2}\;\textrm{Re}b(i \lambda_Y\psi_{\bf b} + h.c.) 
+3\sqrt{2}\;\frac{c_{YB}}{M_{St}} (i \lambda_Y\psi_{\bf b} + h.c.) - 
6\sqrt{2}\;c_{YB}\textrm{Re}b^2 
+\notag\\
& 
6\sqrt{2}\;\frac{c_{YB}}{M_{St}}\;\textrm{Re}b\;g_B 
\big( B_S S^\dagger S + B_{H_1} H_1^\dagger H_1 + B_{H_2} H_2^\dagger H_2
+ B_{D_R}\tilde{D}_R^\dagger\tilde{D}_R + B_{U_R}\tilde{U}_R^\dagger\tilde{U}_R +
\notag\\
&B_{Q}\tilde{Q}^\dagger\tilde{Q} + B_{R}\tilde{R}^\dagger\tilde{R} +
B_{L} \tilde{L}^\dagger \tilde{L}\big) - 6 \frac{c_Y}{M_{St}} 
\left(\sqrt{2} + 2\frac{c_B}{M_{St}}\;\textrm{Re}b_B\right) 
(i \lambda_Y\psi_{\bf b}+ h.c.) +
\notag\\
&2 g_B \left(1+\sqrt{2}\frac{c_B}{M_{St}}\textrm{Re}b \right)
( -3 H_1^\dagger H_1 + 3 B_{H_2} H_2^\dagger H_2
+ 2 \tilde{D}_R^\dagger\tilde{D}_R - 4 \tilde{U}_R^\dagger\tilde{U}_R +
\tilde{Q}^\dagger\tilde{Q} + 6 \tilde{R}^\dagger\tilde{R} - 
3 \tilde{L}^\dagger \tilde{L})  
\Big\},
\label{eq:Dfieldeom0alpha}
\end{align} 
showing that their on-shell expressions are characterized
by the appearance of the saxion field in a non-polynomial form.
\begin{figure}[h!]
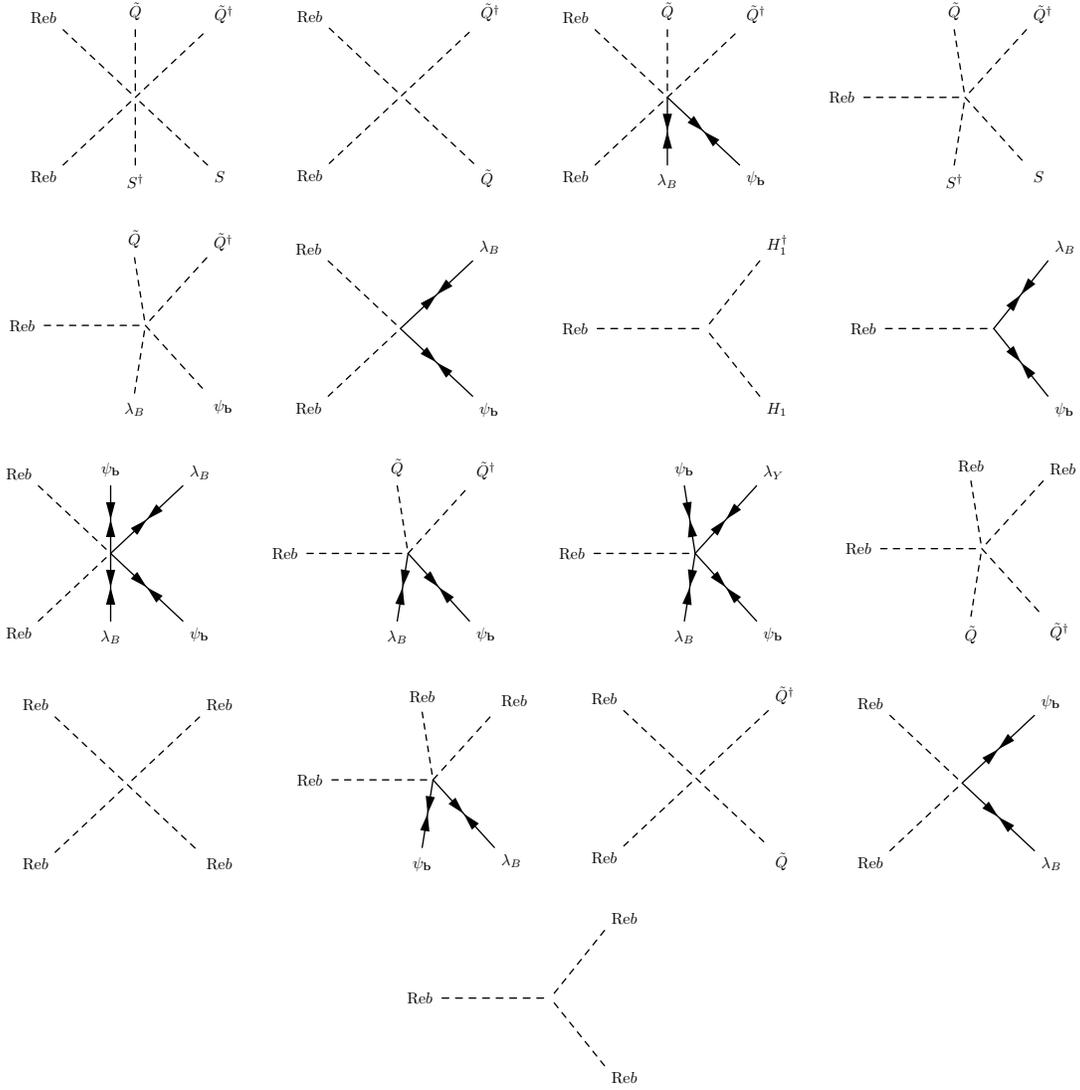
  \centering
\includegraphics[scale=.6]{VerticiReb1.epsi}\hspace{.7cm}
\includegraphics[scale=.6]{VerticiReb2.epsi}\hspace{.7cm}
\includegraphics[scale=.6]{VerticiReb3.epsi}\hspace{.7cm}
\includegraphics[scale=.6]{VerticiReb4.epsi}\\
\vspace{.5cm}
\includegraphics[scale=.6]{VerticiReb5.epsi}\hspace{.7cm}
\includegraphics[scale=.6]{VerticiReb6.epsi}\hspace{.7cm}
\includegraphics[scale=.6]{VerticiReb7.epsi}\hspace{.7cm}
\includegraphics[scale=.6]{VerticiReb8.epsi}\\
\vspace{.5cm}
\includegraphics[scale=.6]{VerticiReb9.epsi}\hspace{.7cm}
\includegraphics[scale=.6]{VerticiReb10.epsi}\hspace{.7cm}
\includegraphics[scale=.6]{VerticiReb11.epsi}\hspace{.7cm}
\includegraphics[scale=.6]{VerticiReb12.epsi}\\
\vspace{.5cm}
\includegraphics[scale=.6]{VerticiReb13.epsi}\hspace{.7cm}
\includegraphics[scale=.6]{VerticiReb14.epsi}\hspace{.7cm}
\includegraphics[scale=.6]{VerticiReb15.epsi}\hspace{.7cm}
\includegraphics[scale=.6]{VerticiReb16.epsi}\\
\vspace{.5cm}
\includegraphics[scale=.6]{VerticiReb17.epsi}\hspace{.5cm}
\caption[small]{Saxion interactions to lowest order in $1/M_{St}$. An
infinite number of additional higher order interactions (in powers of
$1/M_{St}$) are generated by the insertion on these vertices of $n$
powers of saxion lines. We use the double line notation for Majorana
particles. }
\label{neutvert1}
\end{figure} 
The presence of the St\"uckelberg mass allows to perform
an expansion of these terms to all orders in $\textrm{Re}\,b/M_{St}$.
The first terms of the series expansion are given by
\begin{align}
&\frac{1}{12+12\sqrt{2}\textrm{Re}b(c_B + c_Y )/M_{St}  -
6 \textrm{Re}b^2(c_{YB}^2 - 4 c_Y c_B)/M_{St}^2} =\notag\\ 
&\frac{1}{12} - 
c_Y\frac{\textrm{Re}b}{6 \sqrt{2}M_{St}} - 
c_B\frac{\textrm{Re}b}{6 \sqrt{2}M_{St}} + 
\frac{1}{6} c_Y^2 \frac{\textrm{Re}b^2}{M_{St}^2} +
\frac{1}{24} c_{YB}^2 \frac{\textrm{Re}b^2}{M_{St}^2} +
\frac{1}{6} c_B^2 \frac{\textrm{Re}b^2}{M_{St}^2} +
\frac{1}{6} c_B c_Y \frac{\textrm{Re}b^2}{M_{St}^2} + O(\textrm{Re}b^2/M_{St}^2)
\end{align}
We present a list of the vertices to leading order in $1/M_{St}$ in Fig.~\ref{neutvert1}. Additional vertices (not shown) come with $n$ insertions of $\textrm{Re}\, b$ and
a suppression by higher powers $(2 \,n)$ of $M_{St}$.
Some considerations are in order concerning the allowed values of $M_{St}$.  
A very large St\"uckelberg mass, in principle, would be sufficient to
guarantee that the effect of reheating - caused by the decay of the
saxion - takes place well above the temperature of nucleosynthesis
(see for instance the discussion in~\cite{Banks:2002sd}) thereby
avoiding the problem of a possible late entropy release at that
time. In this case one can essentially neglect the saxion from the low
energy spectrum. For moduli of string origin the required mass value
($\sim 100$ TeV), much larger than in our case, is justified by the
suppressed gravitational interaction of the modulus with the rest of
the matter fields and works as an enhancing factor for its decay. In
our case, instead, such a suppression is absent and a fast decay of
the saxion is guaranteed already by a St\"uckelberg mass around the
TeV scale. 

\section{The scalar potential and the saxion} 
As we have mentioned, in
this model there are three scalar fields which take a vev, $H_1, H_2$
and $S$, the scalar components of the scalar superfield $\hat{S}$.
We assume that also the saxion field gets a vev, $v_b$. The scalar potential is composed by contributions coming from the $D$-terms,
$F$-terms and scalar mass terms. Expanding up to $O(\textrm{Re} b/M_{St})$ we find 

\begin{align} 
&V=\vert\lambda
H_{1}\cdot H_{2}\vert ^{2} + \vert\lambda S\vert ^{2}(\vert
H_{1}\vert^{2}+\vert H_{2}\vert^{2}) + m_{1}^{2}\vert H_{1}\vert^{2} + m_{2}^{2}\vert H_{2}\vert^{2} +
m_{S}^{2}\vert S\vert^{2} + (a_{\lambda}SH_{1}\cdot H_{2}+h.c.) -\notag\\
&
 \frac{1}{8}g_Y^2 
\left(H_1^\dagger H_1 - H_2^\dagger H_2\right)^2-\frac{1}{2}g_B^2 
\left(B_{H_1} H_1^\dagger H_1 + B_{H_2} H_2^\dagger H_2  + B_S S^\dagger S\right)^2  -
\notag\\
&
\frac{1}{8} g_2^2 \left[ \left(H_1^\dagger H_1 - H_2^\dagger H_2\right)^2 + 
4 \vert H_1^\dagger H_2 \vert^2 \right]
+\notag\\
&\textrm{Re}b^3 \left[\frac{c_B M_{St}}{\sqrt{2}} 
+ g_Y\frac{\left(c_Y + c_B\right) c_{YB}}{2 M_{St}}
\left(H_1^\dagger H_1 - H_2^\dagger H_2\right) \right] -\notag\\
&\textrm{Re}b^2\bigg[ \frac{m_{\textrm{Re}b}^2}{2}
+ \frac{M_{St}^2}{2} 
+ \sqrt{2}\;c_B\;g_B 
\left(B_{H_1} H_1^\dagger H_1 + B_{H_2} H_2^\dagger H_2 + B_{S} S^\dagger S\right)+ 
\frac{c_{YB}\;g_B}{2\sqrt{2}} 
\left(H_1^\dagger H_1 - H_2^\dagger H_2\right)\bigg]+ 
\notag\\
& \textrm{Re}b\bigg\{
g_B M_{St} \left(B_{H_1} H_1^\dagger H_1 + B_{H_2} H_2^\dagger H_2 +
B_S S^\dagger S \right) + 
\frac{c_W g_2}{32\sqrt{2}M_{St}}
\left[ \left(H_1^\dagger H_1 - H_2^\dagger H_2\right)^2 + 
4 \vert H_1^\dagger H_2 \vert^2 \right] +
\notag\\
&   
\frac{c_Y g_Y^2}{4\sqrt{2}M_{St}} \left(H_1^\dagger H_1 - H_2^\dagger H_2\right)^2 +
\frac{c_B g_B^2}{\sqrt{2}M_{St}} \left(B_{H_1} H_1^\dagger H_1 + B_{H_2} H_2^\dagger H_2  + B_S S^\dagger S\right)^2 
+\notag\\
&
\frac{c_{YB}\,g_Y g_B}{2\sqrt{2}M_{St}}
\left(B_{H_1} H_1^\dagger H_1 + B_{H_2} H_2^\dagger H_2  + B_S S^\dagger S\right)
\left(H_1^\dagger H_1 - H_2^\dagger H_2 \right)
\bigg\}. 
\label{potpot}
\end{align} 

To proceed with the analysis of this potential we introduce the following parameterizations
\begin{equation} H_{1}=\frac{1}{\sqrt{2}}
\begin{pmatrix} \textrm{Re}H_1^{0} + i~\textrm{Im}H_1^{0}\\
\textrm{Re}H_1^{-} + i~\textrm{Im}H_1^{-}
\end{pmatrix} \hspace{0.5cm} H_{2}=\frac{1}{\sqrt{2}}
\begin{pmatrix} \textrm{Re}H_2^{+} + i~\textrm{Im}H_2^{+}\\
\textrm{Re}H_2^{0} + i~\textrm{Im}H_2^{0}
\end{pmatrix} \hspace{0.5cm} S=\frac{1}{\sqrt{2}}(\textrm{Re}S +
i~\textrm{Im}S).  \nonumber\\
\end{equation} expanded around the vevs of the Higgs fields and of the saxion as 
\begin{align} \left\langle H_{1}\right\rangle=
\frac{1}{\sqrt{2}}\left(
\begin{array}{c} v_{1}\\ 0
\end{array} \right) 
\hspace{1cm} 
\left\langle
H_{2}\right\rangle=\frac{1}{\sqrt{2}}\left(
\begin{array}{c} 0\\ v_{2}
\end{array} \right) 
\hspace{1cm}
\tan\beta=\frac{v_2}{v_1}
\hspace{1cm}
\left\langle
S\right\rangle=\frac{v_{S}}{\sqrt{2}}
\hspace{1cm} 
\left\langle
\textrm{Re}b\right\rangle=\frac{v_{b}}{\sqrt{2}}.
\end{align}

The scalar mass parameters can be expressed in terms of the remaining 
parameters of the theory using the minimization conditions for the scalar 
potential. In particular, taking a derivative of the potential with respect to the saxion field we get the relation
\begin{align}
\frac{\partial V}{\partial \textrm{Re}b}=
- v_b m_{\textrm{Re}b}^2 - v_b M_{St}^2  + \frac{1}{2} g_B M_{St} B_{H_1} v_1^2  + 
\frac{1}{2} g_B M_{St} B_{H_2}  v_2^2 + 
\frac{1}{2} g_B M_{St} B_S  v_S^2 
\end{align}
where we have neglected all the contributions suppressed by the St\"uckelberg mass. We can use this relation as a necessary condition in order to express $m_{\textrm{Re}b}^2$ in terms of the vevs and of the other parameters of the scalar potential. A numerical analysis of the Hessian at this point, for the selected parameters of the model used in our simulations, shows that indeed this extremal point indeed corresponds to  a minimum.  

We will try to highlight the most interesting features of
these types of models and the implications for the axion, which are
all connected to the properties of the vacuum of these theories below the scale of SUSY
breaking and at the scales of the electroweak and QCD phase transitions.

\section{Saxion decay modes}
\label{saxion_decay} 
Having summarized the basic structure of the model, we now turn to describe the leading contributions
to the 2-body decays of the saxion
. The goal of this analysis is to ensure that the decay rate of the saxion is such that 
it occurs fast enough in order not to  interfere with the nucleosynthesys. 

We will compute its decay rate by considering the worst possible
scenario, i.e.\ by assuming that this decay occurs around the SUSY
breaking scale, or temperature T around 1 TeV. At this temperature,
the decays of the saxion are parameterized by the typical SUSY
breaking scales such as $M_b$ and $M_Y$, both of O($M_{susy}$). The
model is in a symmetric electroweak phase ($M_{susy} > v$), which
justifies the use of the interaction eigenstates (rather than the mass
eigenstates) for the description of the final decay products.

The relevant interactions for the saxion decay are described by the general Lagrangian
\begin{align} 
{\mathcal L}_{saxion dec.} = & \textrm{Re}\,b \left[
g_B M_{St}~B_{H_1} H_1^\dagger H_1 +
g_B M_{St}~B_{H_2} H_2^\dagger H_2 +
g_B M_{St}~B_S S^\dagger S +
\right.\displaybreak[0]\nonumber\\ 
&\left.  
g_B M_{St}~B_{Q} \sum_{j=1}^{3} \tilde{Q}_{j}^{\dagger} \tilde{Q}_j +
g_B M_{St}~B_{D_R} \sum_{j=1}^{3} \tilde{D}_{R,j}^{\dagger} \tilde{D}_{R,j} +
g_B M_{St}~B_{U_R} \sum_{j=1}^{3} \tilde{U}_{R,j}^{\dagger} \tilde{U}_{R,j}
+\right.\displaybreak[0]
\nonumber\\ 
&\left.
g_B M_{St}~B_{L} \sum_{l=e,\mu,\tau} \tilde{L}_{l}^{\dagger} \tilde{L}_l +
g_B M_{St}~B_{R} \sum_{l=e,\mu,\tau} \tilde{R}_{l}^{\dagger} \tilde{R}_l
\right.\displaybreak[0]\nonumber\\ 
&\left.  
- i\frac{c_B}{2\sqrt{2}}
\left(\lambda_B~\psi_{\bf b} + \bar{\lambda}_B~\bar{\psi}_{\bf
b}\right) - i \frac{c_{YB}}{2\sqrt{2}} \left(\psi_{\bf b}~\lambda_Y+
\bar{\psi}_{\bf b}~\bar{\lambda}_{Y}\right) \right].
\end{align} 
They involve
CP-even and CP-odd massless scalars, the extra singlet scalar $S$, the
squarks, the sleptons and the gauginos $\psi_{\bf b},~\lambda_Y$. We
compute the total decay rate into fermions, squarks and sleptons, and
Higgs scalars.
The left-handed doublets of the squarks and the
sleptons are defined as $\tilde{Q}_j$ and $\tilde{L}_l$ respectively,
while the right-handed singlets are $\tilde{U}_{R,j}$,
$\tilde{D}_{R,j}$ and $\tilde{R}_l$, with $j,l$
labeling the fermion families. 
\begin{itemize}
\item {\bf Decays into fermions}
\end{itemize} Assuming that $M_{\bf b}\approx M_Y$ are slightly less
than 1 TeV, the decay rates of the saxion into one gaugino and one
axino are
\begin{align} &\Gamma\left(\textrm{Re}\,b \rightarrow \bar{\lambda}_B
\psi_{\bf b}\right)= c_B^2\frac{M_{\textrm{Re}\,b}}{32 \pi} \left(1 -
4 \frac{M_{\bf b}^2}{M_{\textrm{Re}b}^2}\right)^{3/2}\,, \nonumber\\
&\Gamma\left(\textrm{Re}\,b \rightarrow \bar{\lambda}_Y \psi_{\bf
b}\right)= c_{YB}^2\frac{M_{\textrm{Re}\,b}}{32 \pi} \left(1 - 4
\frac{M_{Y}^2}{M_{\textrm{Re}b}^2}\right)^{3/2}
\end{align} 
with the expressions of the coefficients $c_B$ and
$c_{YB}$ determining the couplings given explicitly in
Eq.~(\ref{coeffc}). 
Notice that these rates are large due to the linear dependence 
on $M_{\textrm{Re}b}(= M_{St})$.

\begin{itemize}
\item {\bf Decays into Squarks and Sleptons}
\end{itemize} 
In this channel we consider, for simplicity, the decay
only into squarks and sleptons of the same type.  Even in this case we
are assuming that the masses of the squarks and of the sleptons are
all equal and slightly below 1 TeV.  The decay rate into the $i$-type
sfermion is given by \ba \Gamma\left(\textrm{Re}b \rightarrow
\tilde{f}_i^\dagger \tilde{f}_i \right) =
\frac{g_i^2\lambda^{1/2}}{16\pi M_{\textrm{Re}_b}^3} \ea where the
kinematic function $\lambda$ is, in general, defined as
$\lambda=\left(M_i^2 + M_j^2 - M_{\textrm{Re}_b}^2\right)^2 - 4M_i^2
M_j^2$ (here with $M_i=M_j$), and the couplings $g_i$, in the various
cases, are defined as \beq g_i = \left\{
\begin{array}{rl} N_c~c_{U_R} & \textrm{R-handed singlet u-type
squark},\\ N_c~c_{D_R} & \textrm{R-handed singlet quark d-type
squark},\\ N_c~c_{Q_L} & \textrm{L-handed doublet squark},\\ c_{R} &
\textrm{R-handed singlet slepton
$\tilde{e},\tilde{\mu},\tilde{\tau}$},\\ c_{L} & \textrm{L-handed
doublet slepton}.
\end{array} \right.  \eeq Here $N_c=3$ is the color factor and the
various couplings are given as \ba &&c_{D_R}=\frac{1}{2}
g_B~M_{St}~B_{D_R},\hspace{0.5 cm} c_{U_R}=\frac{1}{2}
g_B~M_{St}~B_{U_R}, \hspace{0.5cm} c_{Q_L} = -\frac{1}{2} g_B
M_{St}~B_{Q}, \nonumber\\ &&c_{L}=-\frac{1}{2} g_B~M_{St}~B_{L},
\hspace{0.5 cm}c_{R}=\frac{1}{2} g_B~M_{St}~B_{R}.  \ea

\begin{itemize}
\item {\bf Decays into massless scalars}
\end{itemize}

The decay rate into particles of the Higgs sector that we denote
generically with $h_i=\textrm{Re}H_1, \textrm{Im}H_1,\dots$ is given
by \ba &&\Gamma\left(\textrm{Re}\,b \rightarrow h_i h_i\right)=
\frac{s_i^2}{32 \pi M_{\textrm{Re}\,b}} \left(1 - 4
\frac{M_{s_i}^2}{M_{\textrm{Re}\,b}^2}\right)^{1/2}\,, \ea where the
couplings $s_i$ are defined as \beq s_i = \left\{
\begin{array}{rl} c_{H_1} & \textrm{$H_1$ Higgs doublet}\\ c_{H_2} &
\textrm{$H_2$ Higgs doublet}\\ c_{S} & \textrm{$S$ Higgs singlet}
\end{array} \right.  \eeq and the coefficients $c_{H_1},c_{H_2},c_{S}$
are
\begin{align} 
c_{H_1}=-\frac{1}{4} g_B~M_{St}~B_{H_1},\hspace{1cm}
c_{H_2}=-\frac{1}{4} g_B~M_{St}~B_{H_2},\hspace{1cm}
c_{S}=-\frac{1}{4}g_B~M_{St}~B_{S}.
\end{align}

The total decay rate is obtained by summing over all the decay modes
\begin{align}
\Gamma_{\textrm{tot}}=\Gamma\left(\textrm{Re}\,b \rightarrow
\bar{\lambda}_{b}^{\prime} \psi_{\bf b}\right)
+\Gamma\left(\textrm{Re}\,b \rightarrow \bar{\lambda}_Y \psi_{\bf
b}\right) +\sum_i\Gamma\left(\textrm{Re}\,b \rightarrow
\tilde{\bar{f}}_i \tilde{f}_i \right) +\sum_i\Gamma\left(\textrm{Re}\,
b \rightarrow h_i h_i\right).  
\end{align} 
All the decay rates depend upon the
value of the extra $U_B(1)$ coupling $g_B$, the St\"uckelberg mass
$M_{St}$ and the SUSY breaking scale $M_{susy}$.

\begin{figure}[t]
\begin{center}
\includegraphics[scale=.33,angle=-90]{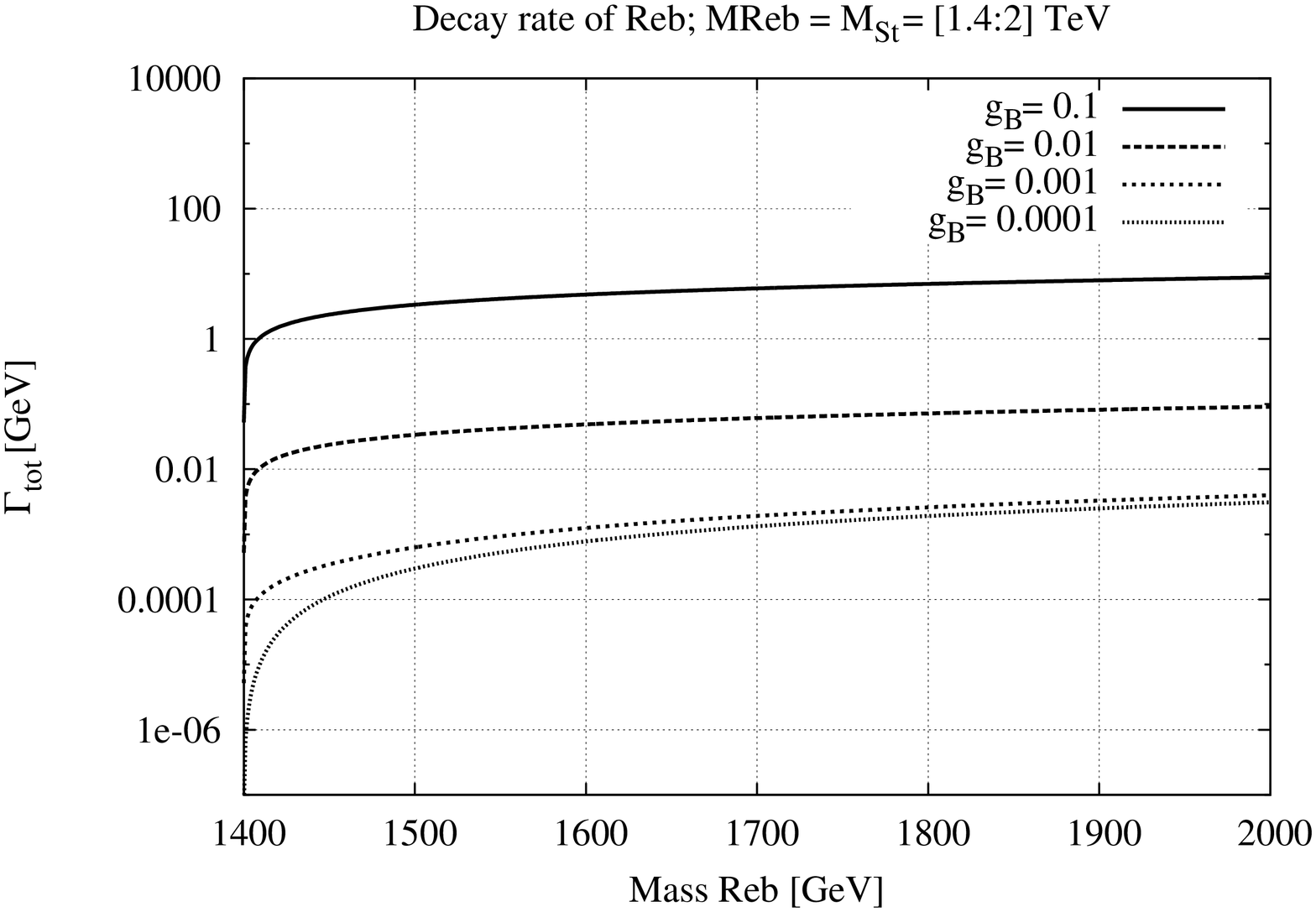}
\includegraphics[scale=.33,angle=-90]{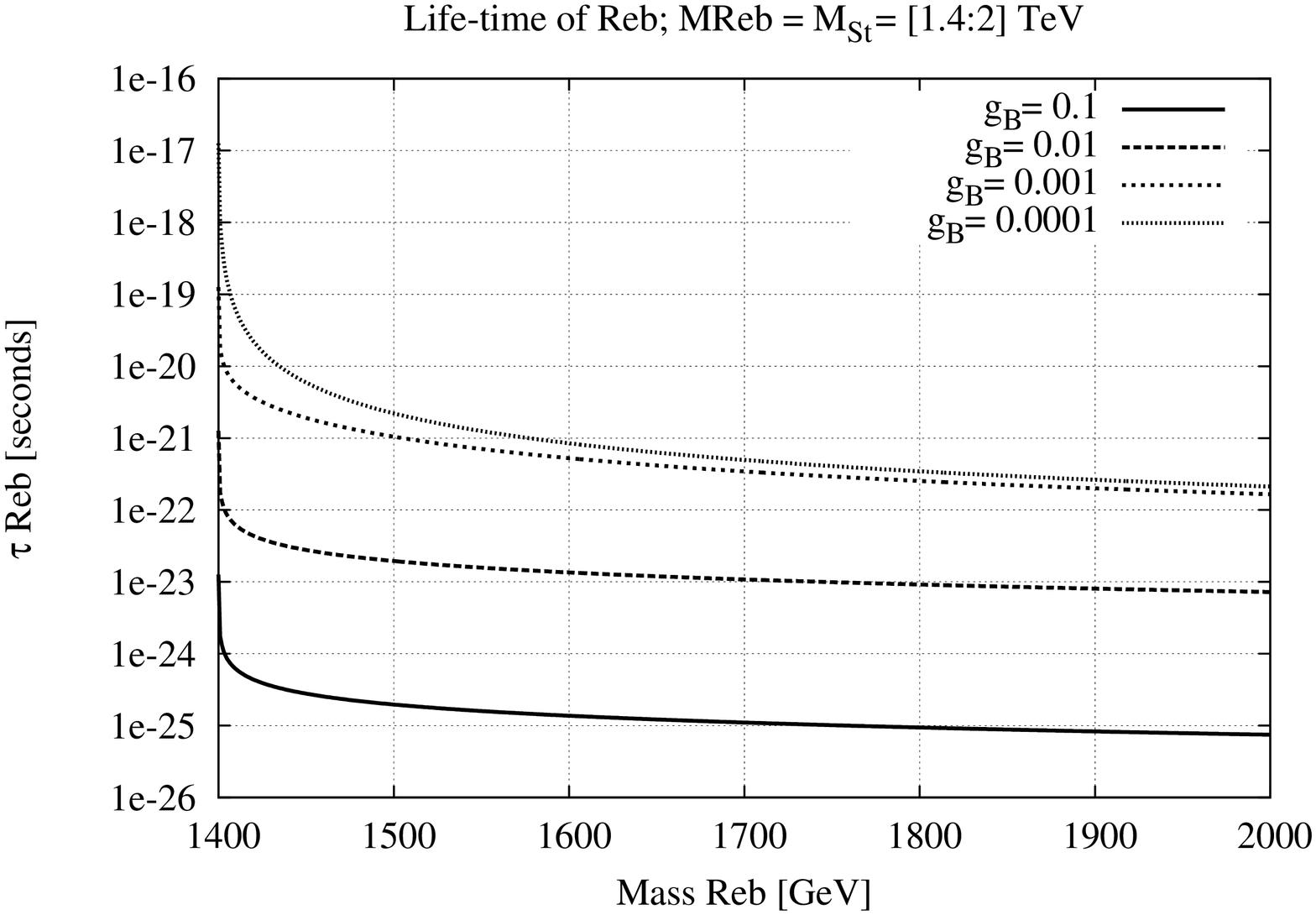}\\
\caption{\small{Total decay rate and lifetime of the saxion for
different values of $g_B$ as a function of the St\"uckelberg mass.}}
\label{decayrate}
\end{center}
\end{figure}

The total decay rate and the lifetime of the saxion are shown in
Fig.~(\ref{decayrate}), with the saxion mass $M_{\textrm{Re}b}$ given by the St\"uckelberg scale ($M_{\textrm{Re}b}=M_{St}$) around 1.4
TeV, and with all the squarks and the sleptons in the final state taken of a mass of 700 GeV. 
All the particles of the Higgs sector are considered to be massless.  For $g_B=0.01$ we obtain a saxion whose decay rate is around 60 MeV if its mass is 1.7 TeV, and which decays
rather quickly, since its lifetime is about $10^{-23}$ seconds. The
lifetime decreases quite significantly as we increase the gauge
coupling of the anomalous gauge symmetry. For instance, for $g_B=0.1$ it decreases to $\sim 10^{-24}$ sec, since the phase space for the decay is considerably enhanced.

 We conclude that the saxion decays sufficiently fast and does not
generate any late entropy release at the time of nucleosynthesis. Obviously, this scenario remains valid for 
all values of the St\"uckelberg mass above the 1 TeV value. 
Therefore, in the analysis of the evolution of the contributions to
the total energy density ($\rho$) of the universe, either due to
matter ($\rho_m$) or to radiation ($\rho_R$), at temperatures $T \leq
2$ TeV, the contribution coming from the saxion is entirely accounted
for by $\rho_R$.

At this point, having cleared the way of any possible obstruction due
to the presence of moduli at the low energy stage ($T \leq 2$ TeV) of
the evolution of our model, we are ready to discuss the relevant
features of the St\"uckelberg field. In particular, we will discuss the appearance of a physical axion, the physical component of the St\"uckelberg, at the electroweak scale. This is extracted from the
CP-odd sector and generated by the mechanism of vacuum
misalignment taking place at the same scale. In particular, the discussion serves to illustrate how
a flat - but physical - direction might be singled out from the vacuum
manifold, acquiring a small curvature at the electroweak phase
transition.

\section{The flat direction of the physical axion from Higgs-axion mixing}
\label{flatsec} 
In \cite{Coriano:2008xa},\cite{Coriano:2008aw} we have
presented in some detail an approximate procedure in order to identify
in the CP-odd sector one state that inherits axion-like interactions.
The approach did not require the explicit expressions of the curvature
terms in the CP-odd part of the supersymmetric potential, which are
instead needed in the discussion of the angle of misalignment.  Here
we are going to extend this analysis by giving the explicit
parameterization of these additional terms. The determination of the
angle of misalignment and its parameterization in terms of the
physical axion is based on an extension of the method presented in
\cite{Coriano:2010py}. We are going to illustrate this point starting,
for simplicity, from the non-supersymmetric case and then moving to
the supersymmetric one.

\subsection{The non-supersymmetric case} 
In the non-supersymmetric
case the scalar sector contains two Higgs doublets $V_{P Q}(H_u, H_d)$ plus one extra contribution (a PQ breaking potential), denoted as $V_{\slashed{P}\slashed{Q}}(H_u,H_d,b)$, 
which mixes the Higgs sector with the St\"uckelberg axion $b$,
\begin{equation}
V=V_{PQ}(H_u,H_d) + V_{\slashed{P}\slashed{Q}}(H_u,H_d,b).
\end{equation}
The mixing induced in the CP-odd sector determines the presence of a linear combination of the St\"uckelberg field $b$ and of the Goldstones of the CP-odd sector, called $\chi$, which is characterized by an almost flat direction, whose curvature is controlled by the strength of the extra potential $V_{\slashed{P}\slashed{Q}}$.  $V_{PQ}(H_u,H_d)$ is the ordinary potential of 2 Higgs doublets,
\begin{align}
V_{PQ}=\mu_u^2 H_u^{\dagger}H_u+\mu_d^2 H_d^{\dagger}H_d+\lambda_{uu}(H_u^{\dagger}H_u)^2
+\lambda_{dd}(H_d^{\dagger}H_d)^2-2 \lambda_{ud}(H_u^{\dagger}H_u)(H_d^{\dagger}H_d)
+2\lambda^{\prime}_{ud}\vert H_u^T \tau_2 H_d\vert^2 \nonumber \\
\end{align}
Concerning the $V_{\slashed{P}\slashed{Q}}$ contribution to the total potential, its
structure is inferred just on the basis of gauge invariance and given by
\begin{align} 
V_{\slashed{P}\slashed{Q}} = &\lambda_0(H_2^{\dagger}H_1 e^{-i g_B
(B_{H_2}-B_{H_1})\frac{b}{2 M_{St}}})+ \lambda_1(H_2^{\dagger}H_1
e^{-i g_B (B_{H_2}-B_{H_1})\frac{b}{2 M_{St}}})^2+\nonumber\\
&\lambda_2(H_2^{\dagger}H_2)(H_2^{\dagger}H_1 e^{-i g_B
(B_{H_2}-B_{H_1})\frac{b}{2M_{St}}})+\lambda_3(H_1^{\dagger}H_1)(H_2^{\dagger}H_1
e^{-i g_B (B_{H_2}-B_{H_1})\frac{b}{2 M_{St}}})+\textrm{h.c.}
\end{align} 
These terms are the only ones allowed by the symmetry of
the model and are parameterized by one dimensionful ($\lambda_0\equiv
\bar{\lambda}_0 v$) and three dimensionless constants
$(\lambda_1,\lambda_2,\lambda_3)$. 

The CP-odd sector is then spanned by the three fields
$(\textrm{Im}H_1,\textrm{Im}H_2, b)$, with the potential $V_{PQ}$ a
function only of $H_1$ and $H_2$. After electroweak symmetry breaking, 
due to Higgs-axion mixing, $b$ can be written as a linear combination of 
a physical axion and of an extra component. The latter is a linear combination 
of the two Goldstone modes of the total potential $(V_{PQ} + V_{\slashed{P}\slashed{Q}})$, 
denoted as $G_0^1, G_0^2$. The physical axion, $\chi$, i.e.\  the component of $b$ 
which is not proportional to the two Goldstones, can be identified using the  
rotation matrix $O^{\chi}$ which relates interaction and mass eigenstates in the 
CP-odd sector
\begin{equation}
\begin{pmatrix} G_0^1 \\ G_0^2 \\ \chi
\end{pmatrix} = O^\chi
\begin{pmatrix} \textrm{Im}H^0_1 \\ \textrm{Im}H^0_2 \\ b
\end{pmatrix}.
\end{equation} 
which takes the form
\begin{equation} b = O_{13}^{\chi} G_0^1 + O_{23}^{\chi} G_0^2 +
O_{33}^{\chi} \chi.
\label{rot12}
\end{equation} 
$\chi$ inherits WZ interactions from $b$ via Eq. (\ref{rot12}), once this is 
introduced into the WZ counterterms.

From an explicit computation one finds that $O_{13}^{\chi}=0$, $O_{23}^{\chi} 
\sim O(1)$ and $O_{33}^{\chi}\sim v/M_{St}$. The Goldstones of the two neutral 
gauge bosons, $G_Z,G_{Z^\prime}$ are linear combinations of $G_0^1$ and $G_0^2$ 
and can be extracted from the bilinear mixings after an expansion around the broken 
electroweak vacuum. Then, the entire CP-odd sector can be spanned by the basis 
$(G_Z,G_{Z^\prime}, \chi)$. The presence of an extra degree of freedom in this sector 
has been established in \cite{Coriano':2005js} using a simple counting of the
degrees of freedom. We review this point for clarity.

There are 9 degrees of freedom in the set $(A_Y, W_3, B, \textrm{Im}
H_1, \textrm{Im} H_2)$, where $B$ is the massive St\"uckelberg gauge vector field, before electroweak symmetry breaking, as well
as 9 in the set $(A_\gamma, Z, Z^\prime, \chi)$, which is generated
after the breaking. The direction determining the gauged axion $\chi$
is then physical but flat, in the absence of an extra potential which
may depend explicitly on $b$. The potential $V^\prime$ is responsible for giving
a small mass for $\chi$ and can be used to parameterize the mechanism
of vacuum misalignment originated at the electroweak scale.

One can explore the structure of this potential and, in particular,
investigate its periodicity. The phase of the potential is indeed
parameterized by the ratio $\chi/\sigma_\chi$ \cite{Coriano:2010py}
\begin{align} 
V_{\slashed{P}\slashed{Q}}=& 4 v_1 v_2 \left(\lambda_2 v_1^2+\lambda_3
v_2^2+\lambda_0\right) \cos\left(\frac{\chi}{\sigma_\chi}\right) + 2
\lambda_1 v_1^2 v_2^2 \cos\left(2\frac{\chi}{\sigma_\chi}\right)
\label{extrap}
\end{align} 
with a mass for the physical axion $\chi$ given by
\begin{align} 
m_{\chi}^2=\frac{2 v_1
v_2}{\sigma^2_\chi}\left(\bar{\lambda}_0 v^2 +\lambda_2 v_1^2
+\lambda_3 v_2^2+4 \lambda_1 v_1 v_2\right) \approx \lambda_{eff} v^2,
\label{axionmass}
\end{align} 
with $\sigma_\chi\sim O(v)$.  The size of this expression is the result of two factors which appear in Eq.~(\ref{axionmass}): the
size of the potential, parameterized by $(\bar{\lambda}_0,
\lambda_1,\lambda_2,\lambda_3)$, and the electroweak vevs of the two
Higgses.  The appearance of $\chi$ in Eq.~(\ref{extrap}) - in the
phase of the extra potential - shows explicitly that the angle of
misalignment is entirely described by this field. The angle is defined
as
\begin{equation} \theta(x)\equiv \frac{\chi(x)}{\sigma_\chi},
\label{angle}
\end{equation} 
where
\begin{equation} 
\sigma_\chi\equiv\frac{2 v_1 v_2 M_{St}}{\sqrt{g_B^2
(B_{H_2}-B_{H_1})^2 v_1^2 v_2^2 +2 M_{St}^2 (v_1^2+v_2^2)}}
\end{equation} is the new dimensionful constant which takes the same role of the scale $f_a$ of the PQ case $(\theta(x)=a/f_a)$. The potential is characterized by a small strength $\sim \lambda_{eff} v^4$, and for this reason one can think of $\chi$ as a pseudo Nambu-Goldstone mode of the theory.

At this stage, it is important to realize that the size of the extra potential is significant in order to establish whether the degree of freedom associated to the axion field remains frozen or not at the electroweak
scale. For instance, if $\lambda_{eff}$ is associated to electroweak
instantons ($\lambda_{eff}\sim \lambda_{inst}$), then $m_\chi$ is very
suppressed (see the discussion in Sec.~\ref{strength}) and far smaller
than the corresponding Hubble rate at the electroweak scale
\begin{equation} 
H(T)=\frac{1}{3}\sqrt{\frac{4}{5}\pi^3
g_{*,T}}\frac{T^2}{M_P}
\end{equation} 
which is about $10^{-5}$ eV.  In the expression above
$g_{*,T_i}$ is the number of effective massless degrees of freedom of
the model at a given temperature ($T$), while $M_P$ denotes the Planck
mass. We recall that the condition \beq m_\chi(T)\sim 3 H(T) \eeq
which ensures the presence of oscillations and determines implicitly
the oscillation temperature $T_i$, is indeed impossible to satisfy if
the misalignment that generates the value of $m_\chi$ at the
electroweak scale is assumed of being of instanton origin (see the
discussion in Appendix~\ref{relics}). This implies that the degree of
freedom associated to this physical axion would be essentially frozen
at the electroweak scale, and the oscillations could take place at a
later stage in the early universe, only around the QCD hadron
transition. Instead, a more sizeable potential, providing an axion
mass larger than $10^{-5}$ eV, would allow such oscillations. For an
axion mass around 1 MeV oscillations indeed occur, but are damped by
the particle decay, given that its lifetime ($\tau_l\sim 10^{-4}$
sec) is much larger than period of their oscillation ( $\tau_{osc}\sim
10^{-13}$ sec). This discussion is going to be expanded to the
supersymmetric case.

\subsection{Supersymmetry and the angle of misalignment } 
In the supersymmetric case the situation is analogous, in the sense that the physical direction $\chi$ can be identified by the same criteria. The superpotential that we are considering allows the presence of one
extra degree of freedom, given by $\text{Im}\,S$, to appear in the
CP-odd sector besides the states $(\textrm{Im}\, H_1, \textrm{Im}\,
H_2, \textrm{Im}\,b)$, already present in the non-supersymmetric case.

From the supersymmetric potential $V$ in~(\ref{potpot}), we identify two massless states, that we
call $G^0_1$ and $G^0_2$, and a massive eigenstate, called $H^0_4$.
$G^0_1$ and $G^0_2$ do not coincide with the true Goldstones of the
model, as in the previous case, since the $V$ potential does not
include any contribution involving $\textrm{Im}\,b$. The correct
neutral Goldstone modes are extracted from the derivative couplings
between the CP-odd scalar fields and the neutral gauge bosons present
in the Lagrangian. The physical axion is then identified as the
massless direction which is orthogonal to the subspace spanned by
$(G_{Z},G_{Z^\prime}, H^0_4)$. This state is called $H_0^5\equiv \chi$
and is given by the linear combination 
\begin{align} 
& \chi=
\frac{1}{N_{\chi}}\left[M_{st}v_1 v_2^2\, \textrm{Im}H^0_1 +
M_{St}v_1^2 v_2\, \textrm{Im}H^0_2 - M_{st}v^2 v_S \,\textrm{Im}\,S -
B_S\,g_B (v^2 v_S^2 + v_1^2 v_2^2)\textrm{Im}\,b\right]
\nonumber\\ 
& N_{\chi}=\sqrt{M_{St}^2 v^2 (v^2 v_S^2 + v_1^2
v_2^2)+B_S^2 g_B^2 (v^2 v_S^2 + v_1^2 v_2^2)^2}.
\label{proj} 
\end{align}
It is important to remark that this state is not constructed, at least
at this stage, from the matrix $O^\chi$, since the projection of
$\textrm{Im}\, b$ on $\chi$ would be zero, if the matrix $O_\chi$ were
derived just from the potential $V$ since it does not depend on
$\textrm{Im}\,b$.\\
Also in this case, the identification of the Goldstones of the two
neutral massive gauge bosons ($Z$ and $Z^{\prime}$) is obtained by
looking at the bilinear mixing terms; these appear in the Lagrangian
once this is rewritten in the physical basis (in the form $M_Z Z\partial
G_Z$, and $ M_{Z^\prime} Z'\partial G_{Z'}$). Then one can immediately 
figure out that the linear basis spanning the
entire CP-odd sector can be completed by the addition of an extra,
orthogonal state $\chi$ $(G_Z, G_{Z^\prime}, H^0_4, \chi)$. The new
entry parameterizes a massless but physical direction in this
sector. Once $\textrm{Im}\, b$ is re-expressed in terms of the
physical axion $\chi$ and of the Goldstone modes $G_Z, G_{Z^\prime}$
of the massive gauge bosons, $\chi$ will inherit from $\textrm{Im} \,
b$ axion-like interactions and will be promoted to a generalized PQ
axion.

At this point, having identified this flat but physical direction of
the potential in the CP-odd sector, one can ask the obvious question
whether the same potential can acquire a curvature. These effects are
indeed parameterized by the strength ($\lambda_{eff}$) of the
potential $V^\prime$ (the ``extra
potential'') that we are going to identify below, and which remains a
free parameter in the theory.

One special comment is deserved by $v_S$, the vev of the
scalar singlet, which is new compared to the standard MSSM scenario
and which is part of the scalar potential.  We recall that this new
scale is essentially bound by the condition $\lambda v_S\sim \mu\sim
10^2-10^3\,\textrm{GeV}$ (see Eq.~\ref{supi}). This defines the typical
range for the $\mu$ term, which sets the scale of the interaction for
the two Higgs doublets in supersymmetric theories.

 In our case we are allowed to parameterize this new non-perturbative
contribution ($V'$) to the potential, as discussed in the previous
section, in a rather straightforward way, by classifying all the
phase-dependent operators which can be constructed using the
fundamental fields of the model. In analogy to the non-supersymmetric
case (the MLSOM) \cite{Coriano':2005js} we rely only on gauge
invariance as a guiding principle to identify them. These include, in
particular, a dependence of $V'$, again in the form of a phase factor, from the
St\"uckelberg field $\textrm{Im}\, b$. 

The contributions appearing in $V'$ don't need to be given
necessarily in a supersymmetric form, since we are assuming that
supersymmetry is already broken at the scale at which they appear ($v
< M_{susy}$). They are parameterized in the form
\begin{equation} 
V'  = \sum_{i=1}^6 V_i
\label{otherpot}
\end{equation} 
where
\begin{align} 
V_1&=a_1 S^4 e^{- i 4 g_B B_S \frac{\mathrm{Im}b}{2
M_{St}}} + h.c.\nonumber\displaybreak[0]\\ 
V_2&=e^{- i g_B B_S \frac{\mathrm{Im}b}{2
M_{St}}}\left(a_2 H_1\cdot H_2 S^2 + b_2 H_1^{\dagger} H_1 S +b_3
H_2^{\dagger} H_2 S +b_4 S^{\dagger} S^2 + d_1 S\right)
+h.c.\nonumber\displaybreak[0]\\ 
V_3&=e^{- i g_B 2 B_S \frac{\mathrm{Im}b}{2
M_{St}}}\left(a_3 H_1^{\dagger} H_1 S^2 + a_4 H_2^{\dagger} H_2 S^2
+a_5 S^{\dagger} S^3 +c_1 S^2 \right) + h.c.\nonumber\displaybreak[0]\\ 
V_4&=a_6
(H_1\cdot H_2)^2 e^{i g_B 2 B_S \frac{\mathrm{Im}b}{2 M_{St}}} +
h.c.\nonumber\displaybreak[0]\\ 
V_5&=b_1 S^3 e^{- i g_B 3 B_S \frac{\mathrm{Im}b}{2
M_{St}}} + h.c.\nonumber\displaybreak[0]\\ 
V_6&=c_2 H_1\cdot H_2 e^{i g_B B_S
\frac{\mathrm{Im}b}{2 M_{St}}} + h.c..
\label{eq:PQcontributions}
\end{align} 
In the expressions above we have grouped together terms
that share the same phase factor. Notice that the parameters $a_i,
b_j$, $c_k$ and $d_1$ carry different mass dimensions. For these
reasons they can be parameterized by suitable powers of the SUSY
breaking mass $M_{susy}$ times $\lambda_{eff}$.  We explicitly obtain
the estimates 
\begin{align} 
a_i \sim \lambda_{eff} \qquad 
b_j \sim \lambda_{eff}\, M_{susy} \qquad 
c_k \sim \lambda_{eff}\, M_{susy}^2 \qquad
d_1 \sim \lambda_{eff}\, M_{susy}^3.
\label{size} 
\end{align}
If we introduce any of the terms in
Eq.~(\ref{eq:PQcontributions}), and recompute the CP-odd mass matrix
using the new potential $(V + V^\prime)$, this gets modified, but we
still find two massless eigenstates corresponding to the neutral
Goldstone modes, which also in this case we call $G_0^1$ and
$G_0^2$. They can be expressed as linear combinations of the neutral
Goldstone states coming from the derivative couplings between the
gauge bosons and the CP-odd Higgs fields. An important point to remark
is that these states (Goldstone modes) do not depend on the parameters
of the Peccei-Quinn breaking potential, as we expect, since the
presence of this extra potential doesn't affect the bilinear
derivative couplings through which they are identified.

In the basis $(\textrm{Im} H_1^1,\textrm{Im} H_2^2,\textrm{Im}
S,\textrm{Im} b)$ they are given by
\begin{align}
G_0^1&=\left\{\frac{v_1}{v},\frac{-v_2}{v},0,0\right\}\nonumber\\
G_0^2&=\frac{1}{\sqrt{M_{St}^2 + 
g_B^2 (B_{H_1}^2 v_1^2 + B_{H_2}^2 v_2^2 + B_S^2 v_S^2) }}
\left\{ g_B B_S \frac{v_1 v_2^2}{v^2},
g_B B_S \frac{v_1^2 v_2}{v^2}, - g_B B_S v_S, M_{St}\right\}.
\end{align} 

\subsection{The strength of the potential and $\lambda_{eff}$}
\label{strength} One important comment concerns the possible size of the axion mass 
$m_\chi$  induced by $V'$ at the electroweak scale.  In this
respect we will take into account two basic possibilities. A first possibility that we will explore 
is to assume that the axion mass is PQ-like, in the milli-eV
region; as a second possibility we will select an axion mass around
the MeV region. These choices cover a region of parameter space that has never been analyzed in these types of models, 
while a study of the GeV region for the axion mass has been addressed before in \cite{Coriano:2009zh}.
These choices have to be confronted with constraints
coming from a) direct axion searches, b) nucleosynthesys constraints and c)
constraints on the relic densities from WMAP data.

A PQ-like axion is bound to emerge in the spectrum of the theory if
the potential $V'$ is strongly suppressed and the real mechanism of
misalignment which determines its mass is the one taking place at the
QCD transition. The value of $\lambda_{eff}$, under these assumptions,
should be truly small and one way to achieve this would be to
attribute its origin to electroweak instantons. Using the
numerical relations for the electromagnetic ($\alpha$) and weak
couplings ($\alpha_W$), $1/\alpha(M_Z)=128$ and $\alpha_W=
\alpha/\sin^2 \theta_W$ with $\sin^2 \theta_W(M_Z)=0.23$ on the $Z$
mass $(\alpha_W(M_Z)=0.034)$, the exponential suppression of the extra
potential is controlled by $\lambda_{eff}\sim e^{-185}\equiv \lambda_{inst}= 4.5\times
10^{-81}$. This corresponds to a mass for the axion given by
$m_\chi\sim \sqrt{\lambda_{eff}} v\sim 10^{-29}$ eV. This mass would
be obviously redefined at the QCD epoch.

As we have briefly mentioned in the introduction, mass values of the
axion field around $10^{-33}$ eV (for global $U(1)$'s or of PQ type)
and with a spontaneous breaking scale $f_a\sim 10^{18}$ eV have been
considered as a possible origin of a cosmological constant
$\Lambda^4\sim \left(10^{-3}\textrm{eV} \right)^4$
\cite{Nomura:2000yk}. In such models the misalignment is purely of
electroweak origin and connected to electroweak
instantons. Oscillations of fields of such a mass would not take
place even at the current cosmological time.
 
 Instead, for an axion of a mass in the MeV region, the value of
$\lambda_{eff}$ is larger $(\sim 10^{-12})$ and will be estimated
below. In this case the effect of vacuum misalignment at the QCD scale
is irrelevant in determining the mass of this particle. A more massive
axion, in fact, decays at a much faster rate than a very light one and
the usual picture typical of a long-lived PQ-like axion, in this specific case, simply does
not apply.

In order to characterize in more detail the potential in
Eq.~(\ref{otherpot}), we proceed with a careful analysis of the field
dependence of the phase factors in the exponentials, that we expect to
be written exclusively in terms of the physical fields of the CP odd
sector, $H_0^4$, and the axion $\chi$ $(\chi\equiv H_0^5)$. In fact,
this is the analogous (and a generalization) of what found in the
previous section (see Eq.~(\ref{extrap})), where the periodicity has
been shown to depend only on the axion $\chi$. For this purpose we use
the following parameterization of the fields
\begin{align} 
H_1^1(x)&=\frac{1}{\sqrt{2}}\left( \rho_1^1(x) + v_1
\right)e^{i\Phi_1^1(x)}\hspace{.5cm}
H_1^2(x)=\frac{1}{\sqrt{2}}\rho_1^2(x) e^{i\Phi_1^2(x)}\nonumber\\
H_2^1(x)&=\frac{1}{\sqrt{2}}\rho_2^1(x)e^{i\Phi_2^1(x)}\hspace{.5cm}
H_2^2(x)=\frac{1}{\sqrt{2}}\left( \rho_2^2(x) + v_2
\right)e^{i\Phi_2^2(x)}\nonumber\\ S(x)&=\frac{1}{\sqrt{2}}\left(
\rho_S(x) + v_S \right)e^{i\Phi_S(x)}
\end{align} 
and select just some of the $V_i$ in
Eq.~(\ref{eq:PQcontributions}) in order to illustrate the general
behaviour.

For instance, if we consider only the $V_1$ term we get the
corresponding symmetric mass matrix for the total potential $V + V_1$,
with $V$ defined in Eq.~(\ref{potpot}),
\begin{align} 
M_{odd}^2=-\frac{a_\lambda}{\sqrt{2}}
\begin{pmatrix} \frac{v_2v_S}{v_1} & v_S & v_2 & 0\\ \\ \cdot &
\frac{v_1v_S}{v_2} & v_1 & 0 \\ \\ \cdot & \cdot & \frac{v_1v_2}{v_S}
+ 8 \sqrt{2} \frac{a_1}{a_\lambda}v_S^2 &
-4\sqrt{2}\frac{a_1}{a_\lambda}\frac{g_B B_Sv_S^3}{M_{St}}\\ \\ \cdot
& \cdot & \cdot & 2\sqrt{2}\frac{a_1}{a_\lambda}\frac{g_B^2 B_S^2
v_S^4}{M_{St}^2}
\end{pmatrix}
\end{align} 
expressed in the basis
$(\Phi_1^1,\Phi_2^2,\Phi_S,\textrm{Im}b)$. From this matrix we get two
null eigenvalues corresponding to the neutral Goldstones and two
eigenvalues which correspond to the masses of the two CP-odd states
$H_0^{4}$ and $H_0^{5}$. In this specific case they take the form
\begin{align} 
m^2_{H_0^{4},H_0^{5}}&=\frac{1}{2
M_{St}v_1v_2v_S}\left(A\pm\sqrt{A^2-B}\right)\nonumber\\ A&=4 a_1 v_1
v_2 v_S^3\left(4 M_{St}^2+g_B^2 B_S^2 v_S^2\right)+\sqrt{2} a_\lambda
M_{St}^2 \left(v_1^2v_2^2+v^2v_S^2\right)\nonumber\\ B&=16 \sqrt{2}
a_1 a_\lambda M_{St}^2 v_1 v_2 v_S^5 \left(4 v^2 M_{St}^2+g_B^2 B_S^2
\left(v_1^2v_2^2+v^2v_S^2\right)\right).
\end{align} In the limit of a vanishing $a_1$ $(\sim \lambda_{eff})$
we obtain a massless state corresponding to $H_0^5$ ($\chi$) and a massive one
corresponding to $H_0^4$. In fact, expanding the expressions above up
to first order in $a_1$, which is a very small parameter due to~(\ref{size}), we obtain for the two eigenvalues the approximate forms
\begin{align} 
&m^2_{H_0^4}\simeq\sqrt{2}a_\lambda
\left(\frac{v_1v_2}{v_S}+\frac{v_1v_S}{v_2}+\frac{v_2v_S}{v_1}\right)+16
a_1\frac{v_1^2v_2^2v_S^2}{v^2v_S^2 + v_1^2v_2^2},\nonumber\\
&m^2_{H_0^5}\simeq\frac{4 a_1 v_S^4 \left[4 v^2 M_{St}^2 + g_B^2 B_S^2
\left(v^2v_S^2+v_1^2v_2^2\right)\right]}{M_{St}^2
\left(v^2v_S^2+v_1^2v_2^2\right)}.
\end{align} 
These relations show that indeed $m_{H_0^5}$ is
$O(\lambda_{eff} v)$ while $m_{H_0^4}$ is $O(v)$.

Moving to the analysis of the phase factor of the same term ($V_1$),
the linear combination of fields that appears in the exponential
factor is given by the expression
\begin{align} \bar\theta_1\equiv\frac{4 \Phi_S(x)}{v_S}-\frac{2 g_B
B_S \textrm{Im}b(x)}{M_{St}}.
\end{align} We rotate this linear combination on the physical basis
$(G_Z,G_{Z},H_0^{4},H_0^{5})$ using the rotation matrix
$O^\chi$. After the rotation we can re-express the angle of
misalignment as a linear combination of the physical states of the
CP-odd sector in the form
\begin{align} \overline{\theta}_1
=\frac{H_0^{4}}{\sigma_{H_0^{4}}}+\frac{H_0^{5}}{\sigma_{H_0^{5}}}.
\label{sigmachi}
\end{align}
This linear combination will appear in all the operatorial terms included in $V'$ and is a generalization of Eq.~(\ref{angle}), with $\sigma_{H_0^{4}}$ and $\sigma_{H_0^{5}}$ defining, separately, the scales of the 
two angular contributions to the total phase. 
\begin{figure}[t]
\centering
\includegraphics[scale=.3, angle=0]{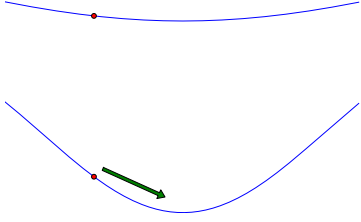}
\caption{\small{Illustration of the two misalignments at the
electroweak (upper figure) and at the QCD phase transitions (lower
figure) for a PQ-like axion (not to scale).}}
\label{misal}
\end{figure} 
 It is not difficult to show that the periodicity of the
potential depends predominantly on $H_0^4$. This can be easily seen by
analyzing the size of $\sigma_{H_0^{4}}$ and
$\sigma_{H_0^{5}}$.  In fact, expanding to first order in $a_1$ we get
\begin{align} 
\frac{1}{\sigma_{H_0^{4}}} &= -\frac{4 v_1
v_2}{v_2\,\textrm{sgn}\,B_S \sqrt{v^2 v_S^2+v_1^2 v_2^2} }-\frac{8
\sqrt{2} v_1^2 v_2^2 v_S^4 \left[4 v^2 M_{St}^2+g_B^2 B_S^2 \left(v^2
v_S^2+v_1^2 v_2^2\right)\right] a_1}{a_\lambda M_{St}^2 \left(v^2
v_S^2+v_1^2 v_2^2\right)^{5/2}
\,\textrm{sgn}\,(B_S)}+O\left(a_1^2\right)\nonumber\\
\frac{1}{\sigma_{H_0^{5\prime}}} &= -\frac{8 \sqrt{2} v_1^2 v_2^2
v_S^4 \left[4 v^2 M_{St}^2+g_B^2 B_S^2 \left(v^2 v_S^2 + v_1^2
v_2^2\right)\right] a_1}{a_\lambda M_{St}^2 \left( v^2 v_S^2 + v_1^2
v_2^2\right)^{5/2}\,\textrm{sgn}\,B_S}+O\left(a_1^2\right).
\end{align} 
with $a_\lambda$ being proportional to the SUSY breaking
scale $M_{susy}$. A more careful look at the structure of these two
scales shows that $\sigma_{H_0^{4}}\sim v_S$ ($v_S= 400$ GeV in our case)
while $\sigma_{H_0^{5}}\sim M_{susy}/\lambda_{eff}$. Clearly,
$\sigma_{H_0^{5}}\gg \sigma_{H_0^{4}}$, but the dependence of the
extra potential on $\chi$ is clearly affected by the different possible sizes of $\lambda_{eff}$. For an instanton generated potential ($\lambda_{eff}\sim \lambda_{inst}$) the direction of $\chi$ is essentially flat and $\sigma_{H_0^{5}}$ turns out to be very large. In turn, this implies that the dependence of the potential $V_1$ on $H_0^5$, which takes place exclusively through the exponential, is negligible, being essentially controlled by $H_0^4$ ($\overline{\theta}_1\sim H_0^4/v$) with 
\begin{align} 
V_1\sim
\lambda_{eff} v^4 \cos(\overline{\theta}_1).
\label{v1pot} 
\end{align}
We may conclude, indeed, that in this case the effect of misalignment 
on $\chi$, generated at the electroweak scale, can be neglected. 
This feature is shown on the left panel of Fig.~\ref{plots12}, where 
we plot $V'(H_0^4,\chi)$. It is immediately clear from these plots that
for $\lambda_{eff}\sim \lambda_{inst}$ the only periodicity of the extra 
potential is in the variable $H_0^4$ (left panel), due to the flatness 
of the $H_0^5$ direction. For a more sizeable potential, with 
$\lambda_{eff}\sim 10^{-12}$, the curvature generated in $\chi$ is 
responsible for giving a mass to the axion in the MeV range (Fig.~\ref{plots12}, right panel). This result is generic for all the terms.
\begin{figure}[t] \centering
\includegraphics[scale=.40]{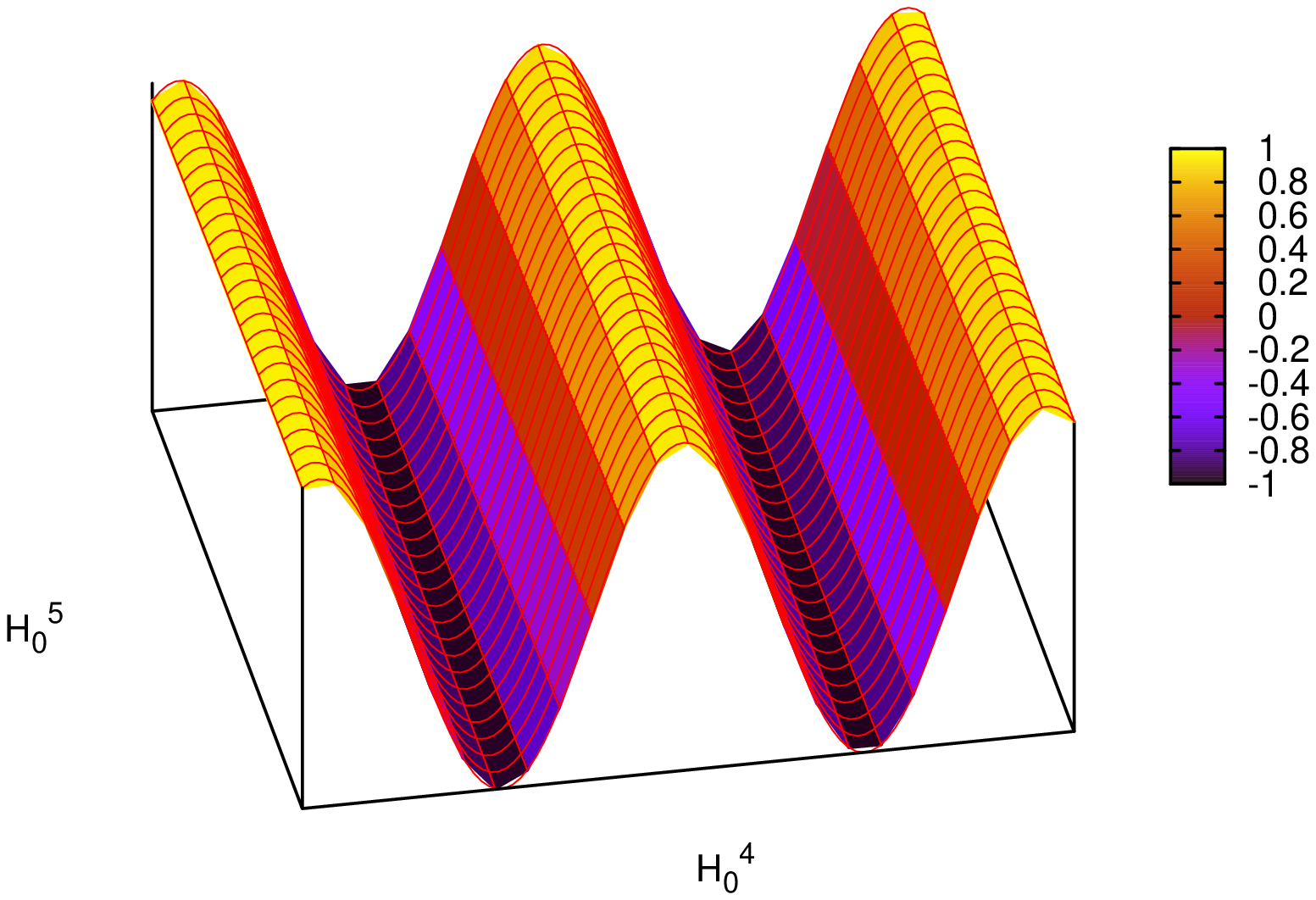}
\includegraphics[scale=.40]{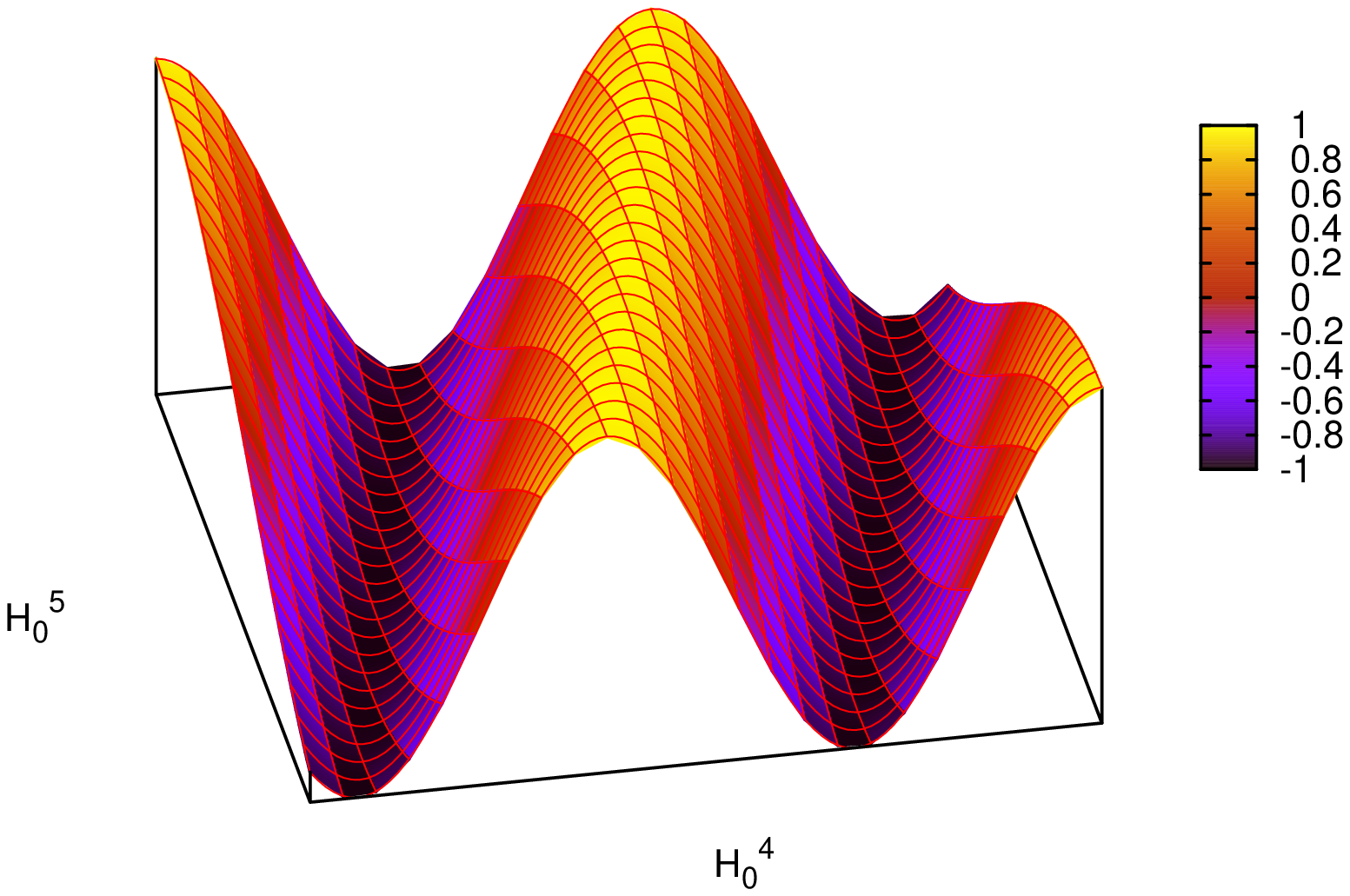}
\caption{Shape of the extra potential $V'$ at the electroweak scale
in the CP-odd sector in the ($H_0^4, \chi\equiv H_0^5$) plane. $\chi$
is an almost flat direction for a strength induced by the instanton
vacuum at the electroweak scale (left panel) and acquires a curvature
for an axion mass in the MeV region (curvature in the $\chi$
direction, right panel).}
\label{plots12}
\end{figure}
One can draw some conclusions regarding the role played by the
exponential phase and compare the supersymmetric with the
non-supersymmetric case.  In the non-supersymmetric case the
periodicity of the potential is controlled by the weak scale ($v$),
and is expressed directly in terms of the physical component of $b$
(which is a real field). The size of the potential, in this case, is
of order $\lambda_{eff} v^4$ \beq V^\prime \sim \lambda_{eff} v^4
\cos\left(\frac{\chi}{v}\right) \eeq and therefore very small, while
the periodicity shows that the amplitude of the axion field is
$\chi\sim O(v)$. In the supersymmetric case, more generally, we obtain
for a generic component $V_i$ \beq V^\prime \sim \lambda M_{susy}^4
\cos\left(\frac{H_0^4}{v_S}+\frac{\chi}{M_{susy}/\lambda_{eff}}
\right)
\label{period} \eeq from which it is clear that the curvature in the
axion field is controlled by the parameter $\lambda_{eff}$. In the
supersymmetric case we can think of the periodicity in
Eq.~(\ref{period}) as essentially controlled by the massive CP-odd
Higgs $H_0^4$, with a period which is $O(\pi v_S)$, with superimposed a
second periodicity of $O( \pi M_{susy}/\lambda_{eff})$ (with
$M_{susy}/\lambda_{eff} \gg v_S$) in the perpendicular direction
($\chi$).  We conclude that the actual structure of the complete ($V +
V^\prime$) potential indeed guarantees the presence in the spectrum of
a physical and light pseudoscalar field. This analysis holds, in
principle, for an axion of any mass, although we do not explicitly
study an axion whose mass goes beyond the MeV region. To have an axion
which is long-lived, the true discriminant of our study is the axion
mass, and for this reason we are going to present a study of the decay
rates of this particle keeping the mass as a free parameter varying in the
milli-eV - MeV interval.

\section{Decay of a gauged supersymmetric axion}
\label{sec:decays-supersymm-gau} 
\begin{figure}[t]
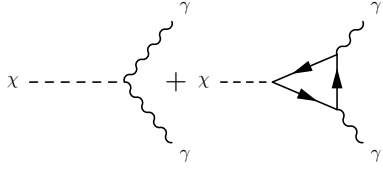
 
\centering
\begin{minipage}[c]{60pt}
\includegraphics[scale=.7]{chigg.epsi}
\end{minipage}+
\begin{minipage}[c]{60pt}
\includegraphics[scale=.7]{chitgg.epsi}
\end{minipage}
\caption{Contributions to the axi-Higgs decay $\chi \,\rightarrow \g
\g$.\label{fig:chi_decay}}
\end{figure}

In this section we compute the decay
rate of the axion of the supersymmetric model into two-photons,
mediated both by the direct PQ interaction and by the fermion loop,
which are shown in Fig.~\ref{fig:chi_decay}, keeping the axion mass as a free parameter.  
 Denoting with $N_c(f)$ the color factor for a fermion
specie, and introducing the function $\tau_f \,\eta(\tau_f)$, a
function of the mass of the fermions circulating in the loop with
\begin{equation} \tau=4m_f^2/m_{\chi}^2\hspace{1cm}
\eta(\tau)=\arctan^2 \frac{1}{\sqrt{- \rho_{f
\chi}^2}}\hspace{1cm}\rho_{f \chi} = \sqrt{1 - \left( \frac{2
m_f}{m_\chi} \right)^2 },
\end{equation} the WZ interaction in Fig.~\ref{fig:chi_decay} is given
by \bea {\mathcal M}^{\mu \nu}_{WZ}(\chi \rightarrow \g \g) = 4
g^{\chi}_{\g\g} \varepsilon[\mu,\nu,k_1,k_2], \eea where
$g^{\chi}_{\g\g}$ is the coupling, defined via the relations
\begin{equation} g^{\chi}_{\g \g}= -\frac{g_B B_S\left(4 c_Y g_2^2 +
c_W g_Y^2\right)}{16 g^2 M_{St}}\sqrt{\frac{v^2 v_S^2+v_1^2 v_2^2}{4
M_{St}^2 v^2 + g_B^2 B_S^2 \left(v^2 v_S^2+v_1^2 v_2^2\right)}}
\end{equation} obtained from the rotation of the WZ vertices on the
physical basis (we will comment in more detail on the size of this
coupling in the next section).  The massless contribution to the decay
rate coming from the WZ counterterm $\chi F^{}_\g F^{}_\g$ is given by
\ba
\label{WZrate1} \Gamma^{}_{WZ}(\chi \rightarrow \g\g)=
\frac{m^3_\chi}{4 \pi}(g^\chi_{\g\g})^2.  \ea Combining also in this
case the tree level decay with the 1-loop amplitude, we obtain for
$\chi \rightarrow \g\g$ the amplitude \ba {\mathcal M}^{\mu \nu}(\chi
\rightarrow \g \g) = {\mathcal M}^{\mu \nu}_{WZ}+{\mathcal M}^{\mu
\nu}_{f}.  \ea

The second amplitude in Fig.~\ref{fig:chi_decay} is mediated by the
triangle loops and is given by the expression
\begin{equation} 
{\mathcal M}^{\mu \nu}_{f}(\chi \rightarrow \g \g) =
\sum_f N_c(f) \,i C_0(m^2_\chi,m_f) c^{\chi, f}_{\g\g}
\varepsilon[\mu,\nu,k_1,k_2] \hspace{.5cm}
f=\{q_u,q_d,\nu_l,l,\chi_1^{\pm},\chi_2^{\pm}\}
\label{pseudo}
\end{equation} 
where $N_c(f)$ is the color factor for the fermions. In
the domain $0< m_\chi < 2 m_f$, which is the relevant domain for our
study, being the axion very light, the pseudoscalar triangle when both
photons are on mass-shell is given by the expression
\begin{equation} 
C_0(m^2_\chi,m_f) = -\frac{m_f}{\pi^2 m_\chi^2}
\arctan^2 \left( \left( \frac{4 m_f^2}{m_\chi^2}
-1\right)^{-1/2}\right) = -\frac{m_f}{\pi^2 m_\chi^2} \eta(\tau)
\label{region1}
\end{equation} 
The coefficient $c^{\chi, f}_{\g\g}$ is the factor for
the vertex between the axi-Higgs and the fermion current. The
expressions of these factors are
\begin{align}
c^{\chi,q_u}&=-\frac{i\,\sqrt{2}\,y_u\,M_{St}\,v_1^2\,v_2}{\sqrt{(v^2
v_S^2+v_1^2v_2^2)\left[4 M_{St}^2 v^2 + g_B B_S (v^2
v_S^2+v_1^2v_2^2)\right]}},\nonumber\\
c^{\chi,q_d}&=-\frac{v_2\,y_d}{v_1\,y_u}\,c^{\chi, q_u},\nonumber\\
c^{\chi,l}&=\frac{y_e}{y_d}\,c^{\chi, q_d}.
\end{align} 
We obtain the following expression for the decay amplitude
\begin{align} 
\Gamma_\chi\equiv\Gamma(\chi \rightarrow \g\g) =&
\frac{m^3_\chi}{32 \pi} \left\{ 8 (g^\chi_{\g\g})^2 + \frac{1}{2}
\left| \sum_f N_c(f) i \frac{\tau_f~\eta(\tau_f)}{4\pi^2 m_f} e^2
Q_f^2 c^{\chi, f} \right|^2 \right.  \nonumber\\ & \left. \qquad\quad
+ 4 g^\chi_{\g\g} \sum_f N_c(f) i \frac{\tau_f~\eta(\tau_f)}{4\pi^2
m_f} e^2 Q_f^2 c^{\chi, f} \right\},
\end{align} where the three terms correspond, respectively, to the
point-like WZ term, to the 1-loop contribution and to their
interference.

Notice that in the expression of this decay rate both the direct
($\sim {(g^\chi_{\g\g}})^2$) and the interference ($\sim
g^\chi_{\g\g}$) contributions are suppressed as inverse powers of the
St\"uckelberg mass, here taken to be equal to 1 TeV. We have chosen
$v$ as the SM electroweak vev, for $v_S$ we have chosen the value
of $500\,\textrm{GeV}$. In order to have an acceptable Higgs spectrum,
the Yukawa couplings have been set to give the right fermion masses of
the Standard Model, while for $g_B$ and $B_S$ we have chosen $g_B=0.1$
and $B_S=4$.

\begin{figure}[t]
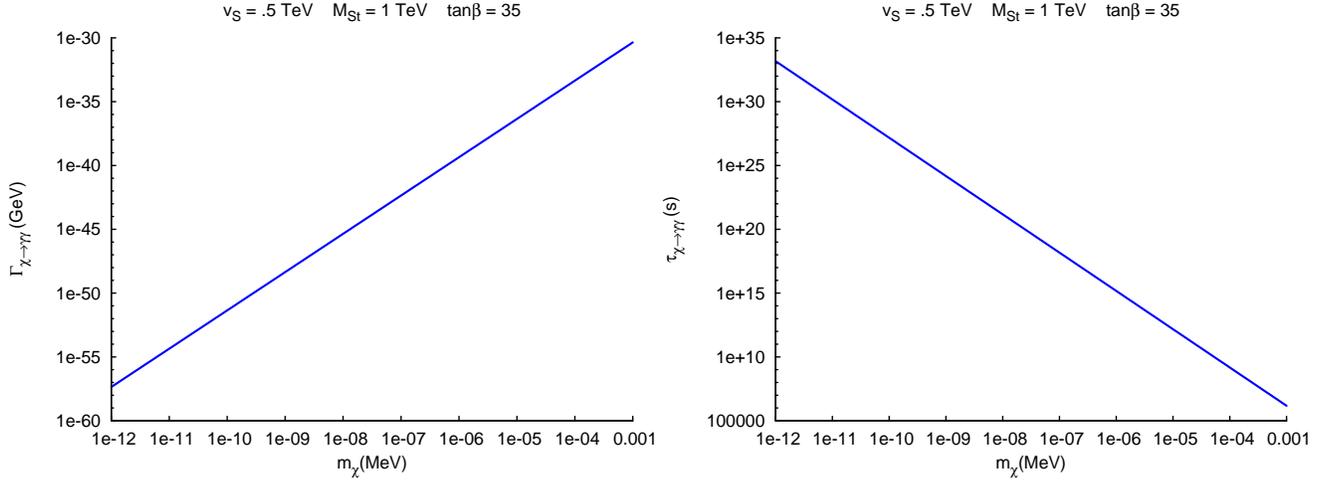
 \centering
\includegraphics[scale=.53]{Gamma_chi.epsi}
\includegraphics[scale=.53]{Tau_chi.epsi}
\caption{Decay amplitude (left panel) and mean lifetime (right panel) for
$\chi\rightarrow\gamma\gamma$ as a function of the axi-Higgs mass.}
\label{fig:chi_decay_plot1}
\end{figure}

\begin{figure}[t]
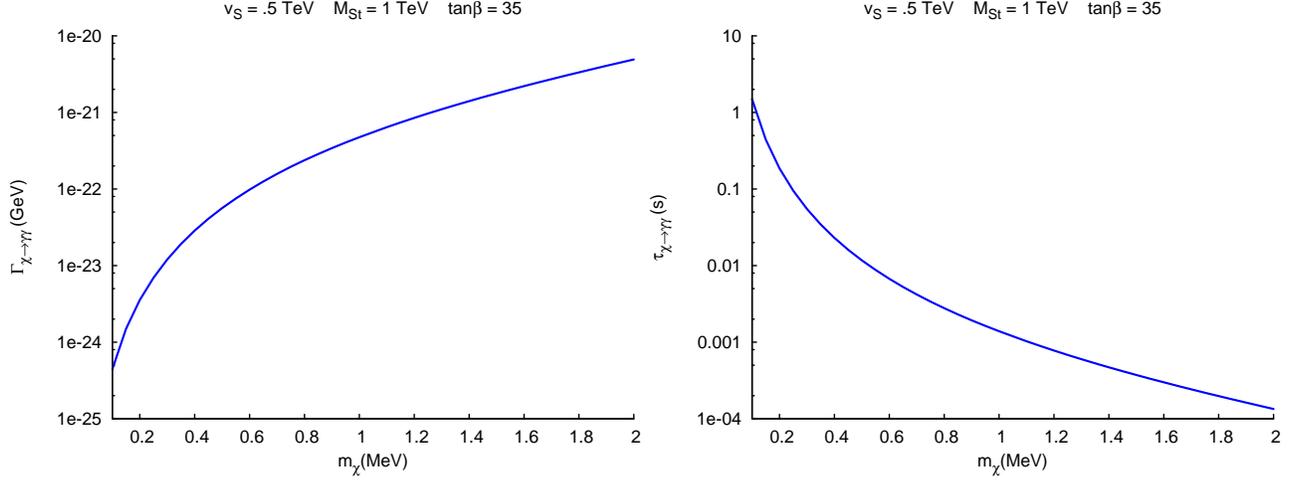
 
\centering
\includegraphics[scale=.53]{Gamma_chimev.epsi}
\includegraphics[scale=.53]{Tau_chimev.epsi}
\caption{Decay amplitude and mean lifetime for
$\chi\rightarrow\gamma\gamma$ as a function of the axi-Higgs mass for
an axion whose mass in the MeV range.}
\label{fig:chi_decay_plot2}
\end{figure}

We show in Figs.~\ref{fig:chi_decay_plot1} and~\ref{fig:chi_decay_plot2} results obtained from the numerical
evaluation of the decay amplitude as a function of the mass of the
axion $m_\chi$, which clearly indicates that the decay rates are very
small for a milli-eV particle, although larger than those of the PQ
case \cite{Coriano:2010py}. We conclude that a PQ-like axion is indeed long-lived also in
these models and as such could, in principle, contribute to the relic densities of dark matter. For an axion with a
mass in the MeV region, instead, the particle is not stable and as such would
decay rather quickly. The decay, in this case, is fast enough
$(\tau\lesssim10^{-3}$ sec) and does not interfere with the
nucleosynthesis.

\section{Cold dark matter by misalignment of the axion field } In the case of a long-lived axion, 
the generation of relic densities of axion dark matter, in this model,
involves two (sequential) misalignments, generated, as we have already
discussed, the first at the electroweak scale, and the second at the
QCD phase transition. The presence of two misalignments at two
separate scales, as discussed in \cite{Coriano:2010py}, is typical of
axions which show interactions both with the weak and with the strong
sectors, due to the presence of mixed anomalies. This point has been addressed in detail within a
non-supersymmetric model, but in a supersymmetric scenario the
physical picture remains the same.

In the PQ-like case, at the first misalignment, taking place at the electroweak scale, the
physical axion is singled out as a component of $\textrm{Im} \,b$,
with a mass which is practically zero, due to the small value of the
curvature induced by the potential generated by electroweak instantons
(Fig.~\ref{misal}, top) given in Eq.~(\ref{otherpot}). In the case of
a very small extra potential ($\lambda_{eff}\sim \lambda_{inst}$) it
is the second misalignment to be responsible for generating an axion
mass. At the second misalignment, taking place at the QCD phase
transition, the mass of this pseudo Nambu-Goldstone mode is redefined
from zero to a small but more significant value ($\sim 10^{-3}$ eV)
induced by the QCD instantons ((Fig.~\ref{misal}, bottom). The final
value of the mass is determined in terms of the hadronic scale
$\Lambda_{QCD}$ and of a second intermediate scale, $M_{St}^2/v$,
which replaces $f_a$ in all of the expressions usually quoted in the
literature and held valid for PQ axions, as we are now going to
clarify. 
\begin{itemize}
\item{\bf MeV axion}
\end{itemize}
An MeV axion is allowed only if the extra potential (the misalignment) is assumed to be generated at a scale different from the 
electroweak phase transition, say at an earlier time. This misalignment, in fact, should be unrelated to the (quasi-periodic) corrections induced at the electroweak time, as shown in Fig. 4, being the latter of very small size.  
However, such an axion would not be long lived. One can easily realize that in this scenario, due to the sizeable value of $m_\chi$, there is an overlap between the
period of coherent oscillations at the QCD hadron transition and the
typical lifetime at which the axion decays. This can be trivially
checked by comparing the QCD time, defined as the inverse Hubble rate
at the temperature of confinement ($H(T_{QCD})\sim 10^{-11}$ eV,
$T_{QCD}\sim 200$ MeV) $t_{QCD}\sim 10^{-4}$ sec with the axion
lifetime in this typical mass range.

\begin{itemize}
\item{\bf PQ-like axion}
\end{itemize}
 For a PQ-like axion the effective scale ($ M_{St}^2/v$) is the result
of the product of two factors: a first factor due to the rotation
matrix of the St\"uckelberg field $\textrm{Im} \,b$ onto $\chi$ -
which is proportional to $v/M$ - times a second factor ($1/M_{St}$)
which is inherited from the original $\textrm{Im} \,b/M_{St} \,
F\tilde{F}$ (WZ) counterterm. Specifically, starting from
Eq.~(\ref{proj}), the size of the projection of $\textrm{Im} \,b$ into
$\chi$ is given by 
\begin{align} 
\frac{1}{N_\chi} B_S\,g_B (v^2 v_S^2 + v_1^2 v_2^2)\sim v/M_{St} 
\end{align}
and hence a typical PQ interaction
term involving the St\"uckelberg field $b$ becomes
\beq \frac{\textrm{Im} \,b}{M_{St}}F F^\prime \to
\frac{\chi}{M_{St}^2/v} F F^\prime. \eeq
The physical state with a $b$ component (i.e. $\chi$) acquires an interaction to $F\tilde F$ which is suppressed by the scale $M_{St}^2/v$.  

Having identified this scale, if we neglect the axion mass generated by
$V'$ (Eq.~\ref{otherpot}) at the electroweak scale, the final mass of
the physical axion induced at the QCD scale is controlled by the ratio
$m_\chi\sim\Lambda_{QCD}^2 v/M_{St}^2$, where the angle of
misalignment is given by $\theta^\prime=\chi v/M_{St}^2$. 

Coming to the value of the abundances for a PQ-like axion - defined as
the number density to entropy ratio $Y=n_\chi/s$ - these can be
computed in terms of the relevant suppression scale appearing in the
$\chi F\tilde{F}$ interaction. We have expanded on the structure of
the computation in Appendix~\ref{relics}. If we indicate with
$\theta^\prime(T_i)$ the angle of misalignment at the QCD hadron
transition and with $T_i$ the initial temperature at the beginning of
the oscillations, the expression of the abundances takes the form
\begin{equation} Y_\chi(T_i)= \left(\frac{v}{M_{St}}\right)\frac{45
M_{St}^2\left(\theta^\prime(T_i)\right)^2}{2\sqrt{5 \pi g_{*, T}} T_i
M_P},
\end{equation} which depends linearly on $M_{St}$. As we have already
mentioned, the computation of the relic densities for a non-thermal
population follows rather closely the approach outlined in the
non-supersymmetric case. For instance, a rather large value of
$M_{St}$, of the order of $10^7$ GeV \cite{Coriano:2010py}, determines
a sizeable contribution of the gauged axion to the relic densities of
cold dark matter. These, in turn, follow rather closely the behaviour
expected in the case of the PQ axion. In practice, to obtain a
sizeable non-thermal populations of gauged axions, $M_{St}$ should be
such that $M_{St}^2/v\sim f_a$, with $f_a$ the usual estimated size of
the PQ axion decay constant. This allows a sizeable contribution of
$\chi$ to the relic density of cold dark matter, with a partial
contribution to $\Omega$ ($\Omega_{\chi}h^2\sim 0.1$) in close analogy to
what expected in the case of the PQ axion. These considerations, which
are in close relations with what found in the non-supersymmetric construction
\cite{Coriano:2010py}, in this case will be subject to the constraints
coming from the neutralino sector and its abundances derived from WMAP. We will come back to this point after presenting the results of 
our simulations in the next sections.

\section{The neutralino sector}
\label{sec:spectr-neutr-sect}
The neutralino sector is constructed
from the eigenstates of the space spanned by the neutral fields
($i\lambda_{W^3}$, $i\lambda_{Y}$, $i\lambda_{B}$,
$\tilde{H_{1}^{1}}$, $\tilde{H_{2}^{2}}$, $\tilde{S}$, $\psi_{\bf
b}$), which involve the three neutral gauginos, the two Higgsinos, the
singlino (the fermion component of the singlet superfield) and the
axino component of the St\"uckelberg supermultiplet. We denote with
$M_{\chi_0}$ the corresponding mass matrix and we list its components
in the appendix.  The neutralino eigenstates of this mass matrix are
labelled as $\tilde\chi_i^0$ ($i=0,\dots,6$) and can be expressed in the basis
$\{i\lambda_{W_3},i\lambda_{Y},i\lambda_{B},
\tilde{H}^1_1,\tilde{H}^2_2,\tilde{S},\psi_{\bf b} \}$ 
\begin{align}
\tilde{\chi}^{0}_i=a_{i1}\,i\lambda_{W_3}+ a_{i2}\,i\lambda_{Y}+
a_{i3}\,i\lambda_{B} + a_{i4}\,\tilde{H}^1_1 + a_{i5}\,\tilde{H}^2_2 +
a_{i6}\,\tilde{S} + a_{i7}\,\psi_{\bf b}.  
\end{align} 
The neutralino mass eigenstates are ordered in mass and the lightest
eigenstate corresponds to $i=0$. We indicate with
$O^{\,{\chi}^0} $ the rotation matrix that diagonalizes the
neutralino mass matrix. In order to perform a numerical analysis of
the model we need to fix some of the parameters, first of all
requiring consistency of their choice with the masses of the Standard
Model particles. For this purpose, the Higgs vev's $v_1$ and $v_2$
have been constrained in order to generate the correct mass values of
the $W^{\pm}$, which depends on $v^2=v_1^2+v_2^2$, and of the $Z$
gauge boson.

The Yukawa couplings have been fixed in order to give the correct masses
of the SM fermions. The choice of $v_S$ and of the parameter $\lambda$
in the trilinear term $\lambda S H_1\cdot H_2$ in the scalar potential
has been made in order to obtain mass values in the Higgs sector in
agreement with the limits from direct searches (with $\lambda \sim
O(1)$). For this reason we have selected the value 
$\lambda=0.5$ and the assignment $B_{H_1} = -1, B_{S} = 3$ for
the $U(1)_B$ charges of the Higgs and the singlet; $B_{Q} = 2$ for the  quark doublet, and $B_{L} = 1$ for the lepton doublet. The gauge mass terms parameters have been selected according to the relation
\begin{align} 
M_Y:M_W:M_G=1:2:6,
\label{scelta0} 
\end{align}
coming from the unification
condition for the gaugino masses. As a further simplification, the sfermion mass parameters
$M_L$, $M_Q$, $m_R$, $m_D$ and $m_U$ have been set to a unique value
$M_0$. We have also chosen a common value $ a_0$ for the trilinear
couplings $a_e$,$a_d$ and $a_u$. With these choices, besides $\tan\beta$, the only other free
parameters left are the St\"uckelberg mass $M_{St}$, the gaugino mass
term for $\lambda_B$, denoted by $M_B$, and the axino mass term,
$M_b$. Our choices are the following
\begin{align} 
&M_Y=500\textrm{GeV} \hspace{.5cm}
M_{YB}=1\, \textrm{TeV}\hspace{.5cm} 
M_W=1 \,\textrm{TeV} \hspace{.5cm} 
M_G=3\, \textrm{TeV}\nonumber\\
&M_L=M_Q=m_R=m_D=m_U=M_0=1\,\textrm{TeV}\nonumber\\
&a_e=a_d=a_u=a_0=1\,\textrm{TeV}\nonumber\\
&a_\lambda=-100\,\textrm{GeV},
\label{scelta1}
\end{align} 
where with $M_L$ and $M_Q$ we have denoted the scalar mass
terms for the sleptons and the squarks, assumed to be equal for all the 3 generations.  
We have also chosen 
\begin{align} 
M_B=M_b=1\,\textrm{TeV}
\label{terza} 
\end{align}
with a coupling constant $g_B$ of the anomalous
$U(1)$ of $0.4$. 
From previous investigations such values of the
anomalous coupling are known to be compatible with LEP data at the $Z$
resonance \cite{Armillis:2008vp, Coriano:2008wf}. In particular, the
mass of the extra $Z^\prime$ ($M_{Z^\prime}$), which in our case is of the order of the St\"uckelberg mass, 
$M_{Z^\prime}\sim M_{St}$ \cite{Armillis:2008vp}, due to the region of variability of $M_{St}$ that we investigate in our simulations, obviously satisfies the current LHC constraints at 95\% CL ( $> 1140$ GeV) from CMS \cite{Chatrchyan:2011wq} and from ATLAS ($ >$ 1.83 TeV) \cite{Collaboration:2011dca} on the absence of a resonance in the dilepton channel at 7 TeV for an extra Z prime with Standard-Model like couplings. This parameter choice is our benchmark, which
is compatible with all the SM requirements on the spectrum of the
known particles. It involves SUSY breaking scales in a kinematical
range which is under investigation at the LHC. We also assume a value $v_b=20$ GeV for the vev of the saxion field Re$b$.

The most significant parameters in the relic density calculation are $M_{St}$ and the Higgs vev ratio $\tan\beta$.
Concerning the St\"uckelberg mass, its value has been chosen to be varied in two different regions, $2-10$ TeV and $11-25$ TeV. In both regions we will consider different values of $\tan\beta$.

\section{Neutralino relic densities and cosmological bounds}
\label{sec:neut_relic_dens} 
As it is well known, the evaluation of the
relic densities requires the calculation of a great number of thermally averaged cross
sections, given the number
of particles which are present. Before coming to the discussion of the
results of this very involved analysis, which is summarized just in
some simple plots of the relic densities of the lightest neutralino -
as a function both of $M_{St}$ and $\tan\beta$, - we present a general description of the structure of the
interactions in the model. We also list the 2-to-2 processes that have
been considered in the coupled Boltzmann equations.

\begin{figure}[t]
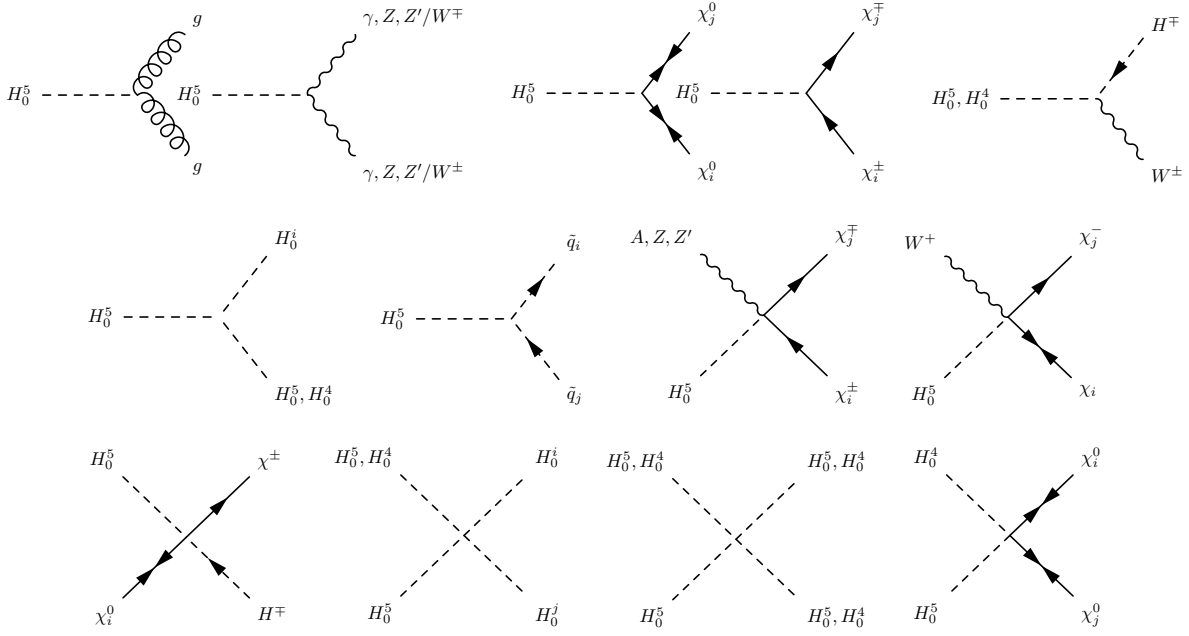
 \centering
\includegraphics[scale=.7]{WZ.epsi}\hspace{.5cm}
\includegraphics[scale=.7]{ah2chi0.epsi}\hspace{.5cm}
\includegraphics[scale=.7]{ahgauge.epsi} \\ \vspace{.5cm}
\includegraphics[scale=.7]{ahcpecpo.epsi}\hspace{.5cm}
\includegraphics[scale=.7]{ahsfer.epsi}\hspace{.5cm}
\includegraphics[scale=.7]{ah2chZ.epsi}\hspace{.5cm}
\includegraphics[scale=.7]{ah2chW.epsi}\hspace{.5cm} \\ \vspace{.5cm}
\includegraphics[scale=.7]{ahchipmchi0Hpm.epsi}\hspace{.5cm}
\includegraphics[scale=.7]{ahcpocpecpe.epsi}\hspace{.5cm}
\includegraphics[scale=.7]{ahcpocpocpo.epsi}\hspace{.5cm}
\includegraphics[scale=.7]{ahH042chi0.epsi}
\caption[small]{Axi-Higgs ($H^5_0$) interactions. The double arrows
denote Majorana particles (neutralinos).}
\label{fig:axionvert}
\end{figure}

\begin{figure}[t]
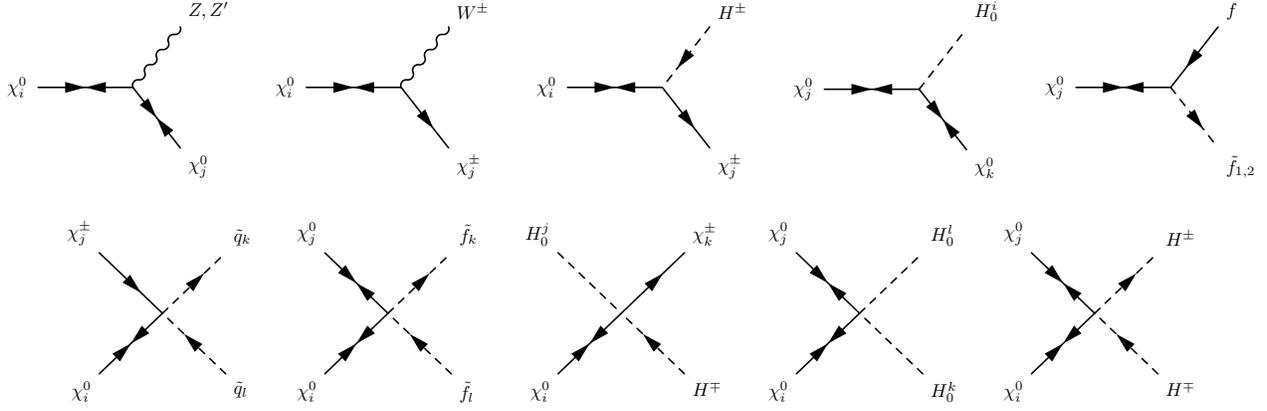
 \centering
\includegraphics[scale=.7]{2chi0gauge.epsi}\hspace{.5cm}
\includegraphics[scale=.7]{chi0chipmgauge.epsi}\hspace{.5cm}
\includegraphics[scale=.7]{chi0chipmHpm.epsi}\hspace{.5cm}
\includegraphics[scale=.7]{2chi0cpev.epsi}\hspace{.5cm}
\includegraphics[scale=.7]{chi0fersfer.epsi} \\ \vspace{.5cm}
\includegraphics[scale=.7]{chi0chipmsfer.epsi}\hspace{.5cm}
\includegraphics[scale=.7]{2chi0sfer.epsi}\hspace{.5cm}
\includegraphics[scale=.7]{chi0chipmH0Hpm.epsi}\hspace{.5cm}
\includegraphics[scale=.7]{2chi02H0.epsi}\hspace{.5cm}
\includegraphics[scale=.7]{2chi02Hpm.epsi}
\caption[small]{Neutralino interactions}
\label{fig:neutvert}
\end{figure}

We start from the action involving the physical axion ($H_0^5$) and
its interactions with the various sectors. These involve, typically,
interactions with the Higgs sector via bilinear vertices (proportional
to $R^{H_0^5 \, H H}$), and trilinear ones (proportional to $R^{H_0^5
\, H H H}$) in $H$, with $H$ denoting generically CP-even and CP-odd
Higgs eigenstates. Other interactions in the same component of the
Lagrangian involve axion-neutralino terms
($R^{H_0^5\,\chi_i^0\chi_j^0}$) plus axion-charginos
($R^{H_0^5\,\chi_i^\pm\chi_j^\mp}$). Other terms are those involving
interactions of the axion with the sleptons $(R^{H_0^5
\tilde{l}_i^{\,\dagger} \tilde{l}_j})$ and the squarks $(R^{H_0^5
\tilde{q}_{i}^\dagger \tilde{q}_{j}})$; vertices involving gauge
bosons (for instance $R^{H_0^5 A \chi^{\pm}_{i} \chi^{\mp}_{j}}$, with
a photon $A$ and two charginos) and quartic contributions with 2, 3
and 4 axion lines.  The Lagrangian describing all the tree-level
interactions involving the axion is
\begin{align} 
{\cal L}_{H_0^5-int}=&R^{H_0^5 H_0^4 H_0^i }H_0^5 H_0^4
H_0^i + R^{H_0^{5\,\,2} H_0^i}\left(H_0^{5}\right)^2 H_0^i + R^{H_0^5
H_0^4 H_0^{i} H_0^{j}}H_0^5 H_0^4 H_0^{i} H_0^{j}+ R^{H_0^{5\,2}
H_0^{i} H_0^{j}} \left(H_0^{5}\right)^2 H_0^{i} H_0^{j}+ \nonumber\\ &
R^{H_0^5 H_0^{4\,3}} H_0^5 H_0^{4\,3}+ R^{H_0^{5\,2} H_0^{4\,2}}
\left(H_0^{5}\right)^2 \left(H_0^{4}\right)^2+ R^{H_0^{5\,3} H_0^{4}}
\left(H_0^{5}\right)^3 H_0^{4}+ R^{H_0^{5\,4}} \left(H_0^{5}\right)^4
+ \nonumber\\ & R^{H_0^5 \chi^0_i \chi^0_j} H_0^5 \chi^0_i \chi^0_j+
R^{H_0^5 \chi^{\pm}_i \chi^{\mp}_j} H_0^5 \chi^{\pm}_i \chi^{\mp}_j+
R^{H_0^5 \tilde{l}_i \tilde{l}_j} H_0^5 \tilde{l}_i^{\,\dagger}
\tilde{l}_j+ R^{H_0^5 \tilde{q}_{i} \tilde{q}_{j}} H_0^5
\tilde{q}_{i}^{\dagger} \tilde{q}_{j}+ \nonumber\\ & R^{H_0^5 A
\chi^{\pm}_{i} \chi^{\mp}_{j}} H_0^5 A^{\mu}
\bar{\chi}^{\pm}_{i}\gamma_{\mu} \chi^{\mp}_{j}+ R^{H_0^5 Z
\chi^{\pm}_{i} \chi^{\mp}_{j}} H_0^5 Z^{\mu}
\bar{\chi}^{\pm}_{i}\gamma_{\mu} \chi^{\mp}_{j}+ R^{H_0^5 Z'
\chi^{\pm}_{i} \chi^{\mp}_{j}} H_0^5 Z^{\prime\mu}
\bar{\chi}^{\pm}_{i}\gamma_{\mu} \chi^{\mp}_{j}+ \nonumber\\ &
R^{H_0^5 W^{\mp} \chi^{\pm}_{i} \chi^{0}_{j}} H_0^5 W^{\mp}_{\mu}
\bar{\chi}^{0}_{j}\gamma^{\mu} \chi^{\mp}_{i}+ R^{\chi^0_i
\chi^{\pm}H^{\mp} H_0^5}\chi^0_i \chi^{\pm}_j H^{\mp} H_0^5+
R^{\chi^0_i \chi^0_j H_0^5 H_0^4}\chi^0_i \chi^0_j H_0^5 H_0^4.
\end{align} 
The explicit expressions of these vertices are rather involved and we omit them. Other interactions
appearing in the interaction Lagrangian involve derivative couplings
with the gauge bosons and the Higgses and they are given by
\begin{align} {\cal L}_{H_0^5-int}&=R^{H_0^5 H^{\pm} W^{\mp}}H_0^5
W_{\mu}^{\mp} \partial^{\mu} H^{\pm}+ R^{H_0^5 H_0^{i} A}H_0^5
A_{\mu} \partial^{\mu} H_0^{i}+ R^{H_0^5 H_0^{i} Z}H_0^5
Z_{\mu} \partial^{\mu} H_0^{i}+ \nonumber\\ & R^{H_0^5 H_0^{i}
Z'}H_0^5 Z_{\mu}^{\prime} \partial^{\mu} H_0^{i}.
\end{align} Similar interactions are also typical for $H_0^4$, the
CP-odd Higgs. Some of the vertices are illustrated in
Fig.~\ref{fig:axionvert}.
 Besides the interaction with the axi-Higgs, we have the
following vertices involving neutralinos
\begin{align} 
{\cal L}_{\chi_0-int}&=R^{\chi^0_i \chi^0_j Z} Z^{\mu}
\bar{\chi}^0_i \gamma_{\mu} \chi^0_j+ R^{\chi^0_i \chi^0_j Z'}
Z^{\prime\mu}\bar{\chi}^0_i \gamma_{\mu} \chi^0_j+ R^{\chi^0_i
\chi^{\pm}_j W^{\mp}} W^{\mp}_{\mu} \bar{\chi}^0_i \gamma_{\mu}
\chi^{\pm}_j+ R^{\chi^0_i \chi^{\pm}_j H^{\mp}} H^{\mp} \chi^0_i
\chi^{\pm}_j+ \nonumber \\ & R^{\chi^0_i \chi^0_j H_0^k} H_0^k
\chi^0_j \chi^0_k+R^{\chi^0_i f \tilde{f}} \chi^0_i f \tilde{f}_{1,2}+
R^{\chi^0_i \chi^{\pm}_j\tilde{q}^{\dagger}\tilde{q}}\chi^0_i
\chi^{\pm}_j \tilde{q}^{\dagger}_k \tilde{q}_l+ R^{\chi^0_i \chi^0_j
\tilde{f}^{\dagger}\tilde{f}}\chi^0_i \chi^{\pm}_j
\tilde{f}^{\dagger}_k \tilde{f}_l+\nonumber \\ & R^{\chi^0_i
\chi^{\pm}_j H^{\mp} H_0^k}\chi^0_i \chi^{\pm}_j H^{\mp} H_0^k+
R^{\chi^0_i \chi^{0}_j H_0^k H_0^l}\chi^0_i \chi^0_j H_0^k H_0^l+
R^{\chi^0_i \chi^{0}_j H^{\pm} H^{\mp}}\chi^0_i \chi^0_j H^{\pm}
H^{\mp}.
\end{align} 
Some of these vertices are illustrated in Fig.~\ref{fig:neutvert}.
The full lagrangian has been implemented using the FeynRules\cite{Christensen:2008py} package. The same package allows to generate the CalcHEP\cite{Pukhov:2004ca} model files which are needed by micrOMEGAs\cite{Belanger:2010gh}
for the calculation of the scattering cross section that are required in the relic density calculation.

With our choice for the parameters the lightest neutralino is the Lightest Supersymmetric Particle (LSP) and so it is the dark matter component in our simulations. The value of the neutralino mass in this case turns out to be 
around 23 GeV with a rather mild dependence on $\tan\beta$. For $\tan\beta$ varying between 5 and 25 the neutralino mass varies from 22.4 to 23.8 GeV. 

We show in Tab.~\ref{tab:neut_ann_chs} a list of
the most relevant 2-to-2 processes which are generated in the $s, t$
and $u$ channels having neutralinos in the initial state (in), while
the possible final states are shown on the right-hand side of the same
table (out).
\begin{table} \centering
\begin{tabular}{ccl} \hline in & s-channel & out\\ \hline
\multirow{4}{*}{$\chi^0_i\,\chi^0_j$ } & $Z$,$Z^{\prime}$ &
$H^{\pm}H^{\mp}$,$H_0^k H_0^4$,$H_0^k H_0^5$,$Z/Z^{\prime}
H_0^k$,$\bar{f}f$,$\tilde{f}^{\dagger}\tilde{f}$\\ & $H_0^k$ &
$H^{\pm}H^{\mp}$,$H_0^l H_0^m$,$H_0^4 H_0^4$,$H_0^4 H_0^5$,$H_0^5
H_0^5$,$Z/Z^{\prime} H_0^4/H_0^5$,\\ &&$W^{\pm} H^{\mp}$,$Z/Z^{\prime}
Z/Z^{\prime}$,$W^{\pm}W^{\mp}$,$\bar{f}f$,$\tilde{f}^{\dagger}\tilde{f}$\\
& $H_0^4,H_0^5$ & $H_0^k H_0^4$,$H_0^k H_0^5$,$Z/Z^{\prime}
H_0^k$,$W^{\pm} H^{\mp}$,$\bar{f}f$,$\tilde{f}^{\dagger}\tilde{f}$\\
\hline \\
\end{tabular}
\begin{tabular}{ccl} \hline in & t/u-channel & out\\ \hline
\multirow{4}{*}{$\chi^0_i\,\chi^0_j$ } & $\chi^0_k$ & $H_0^l
H_0^m$,$H_0^l H_0^4/H_0^5$,$H_0^4/H_0^5 H_0^4/H_0^5$,\\
&&$Z/Z^{\prime}H_0^l/H_0^4/H_0^5$,$Z/Z^{\prime} Z/Z^{\prime}$\\ &
$\chi^{\pm}_k$ & $W^{\pm}/H^{\pm} W^{\mp}/H^{\mp}$\\ & $\tilde{f}$ &
$\bar{f}f$\\ \hline
\end{tabular}
\caption{Tree level neutralino annihilation processes in the 3
kinematic channels}
\label{tab:neut_ann_chs}
\end{table}

\begin{figure}[t]
\centering
\includegraphics[scale=.6]{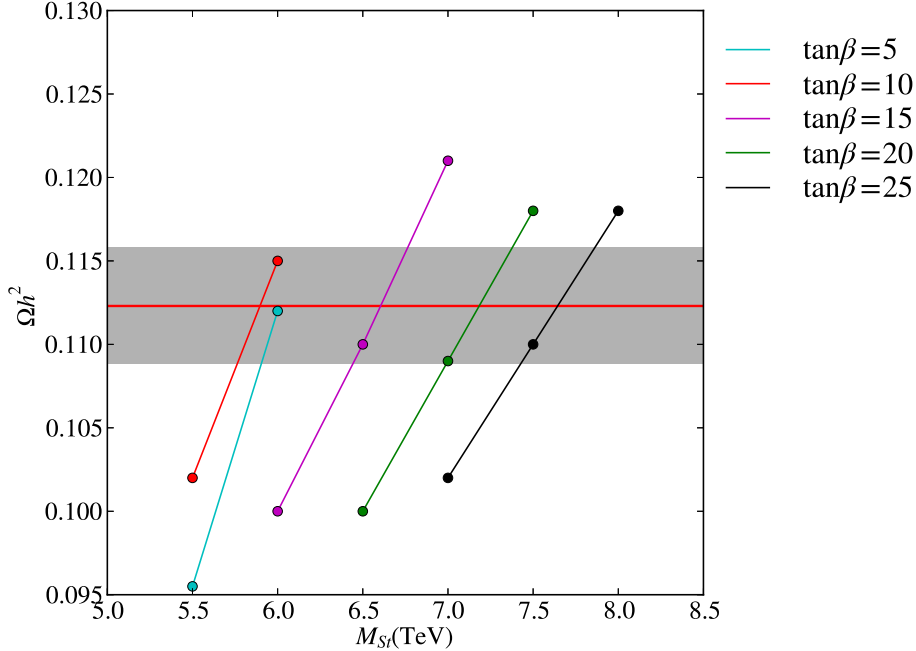}
\caption{Relic density of the lightest neutralino as a function of the St\"uckelberg mass in the region $5-8.5$ TeV for different values of the Higgs vev ratio $\tan\beta$}
\label{fig:lowMst_tanb_relic}
\end{figure}

\begin{figure}[t]
\centering
\includegraphics[scale=.6]{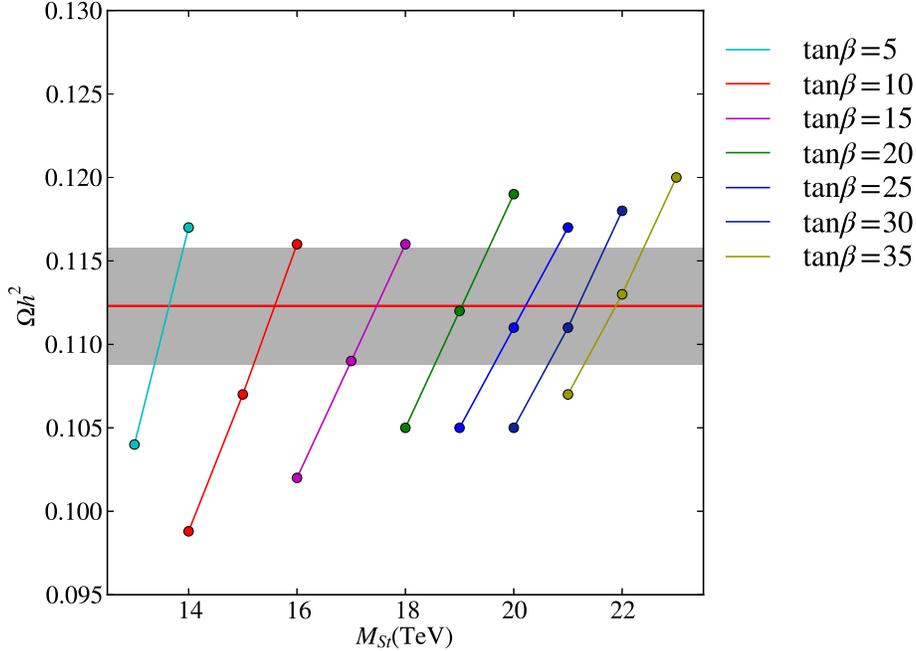}
\caption{Relic density of the lightest neutralino as a function of the St\"uckelberg mass in the region $2-24$ TeV for different values of the Higgs vev ratio $\tan\beta$}
\label{fig:highMst_tanb_relic}
\end{figure}

In Fig.~\ref{fig:lowMst_tanb_relic} we show the results obtained for the lightest neutralino relic density with $M_{St}$ in the range $5-8$ TeV, $v_S=600$ GeV and a varying $\tan\beta$. The values of $v_S$, $\tan\beta$ and $M_{St}$ for which we plot the result coming from the relic density calculation are those that give also acceptable mass values for the whole spectrum, in particular for the neutral Higgs ($\sim 124-126$ GeV) \cite{Chatrchyan:2012tx,ATLAS:2012ae}. The horizontal bar represents the experimental value for the physical dark matter density measured by WMAP, $\Omega h^2=.1123\pm.0035$\cite{Jarosik:2010iu}.
In Fig.~\ref{fig:highMst_tanb_relic} we show the analogous results obtained in the range $11-24$ TeV with $v_S=1.2$ TeV and varying $\tan\beta$. Once again these values are such that we obtain acceptable values for the masses of all the particles in the model. One can immediately notice that for a fixed value of $\tan\beta$ as we increase $M_{St}$, the relic densities grow and tend to violate the WMAP bound. This trend has been found over a sizable range of variability of 
$\tan\beta$ and is a central feature of the model. It is then obvious, from the same figures, that it is possible to raise the 
Stuckelberg mass and stay below the bound if, at the same time, we increase $\tan\beta$.

\section{Summary: windows on the axion mass} 
At this point, before
coming to our conclusions, we can try to gather all the information that
we have obtained so far in the previous sections, summarizing the basic properties of axions in these types of models.
\begin{itemize}
\item{\bf The milli-eV (PQ-like) axion}
\end{itemize} One possibility that we have explored in this work is
that $V'$, the extra potential which is periodic in the axion field,
may be generated around the TeV scale or at the electroweak phase
transition. The actual strength of the potential, remains, in our
construction, undetermined and the physical features of the axion
(primarily its mass), depend upon this parameter. We have tried to
describe the various possibilities, in this respect, and the essential
features for each choice for the value of the mass. In particular, if the
extra potential is generated by non-perturbative effects at the
electroweak phase transition, then the mass of the axion is tiny and the true mechanism of misalignment which determines its mass
takes place at a second stage, at the QCD phase transition.  In this
case the physical axion of the model would be no much different from
an ordinary PQ axion and would be rather long-lived. At the same time, its abundances are fixed by the possible value of the 
scale $M_{St}^2/v$, which should be rather large ($\sim 10^{10}-10^{12}$ GeV), of the same order of $f_a$ in typical axion models, to be a significant component of cold dark matter.  In the region that we have analyzed numerically, with 
$M_{St}$ around the 2-20 TeV's, the contribution to dark matter from misalignment of the axion field, in this case, should be small.  

A second important constraint on this particle, in this mass range, comes from direct axion searches, which also requires 
the interaction of the axions with the gauge fields (in particular the photon) to be suppressed by a large $f_a$. 
For this reason, with $M_{St}$ in the TeV region, these simulations indicate that  an axion of this mass, 
in fact, can be excluded by typical searches with detectors of Sikivie type. The reason is rather obvious, since
axions in the milli-eV mass range could be copiously produced at the center of the sun and probably should have been seen by now in ground based detectors (helioscopes), such as CAST \cite{Ribas:2009pr}. We recall that one of the goals of searches with helioscopes is to set a lower bound on
the suppression scale $f_a$ of the axion-photon vertex, which is
currently experimentally constrained, as we have already mentioned, to
be rather large.

\begin{itemize}
\item{\bf The meV axion}
\end{itemize} A second possibility that we have investigated is that
the extra potential appearing in the CP-odd sector is unrelated to
instanton corrections in the electroweak vacuum. In this case the mass
of the axion remains a free parameter. The range that we have explored
in this second case involves an axion mass in the MeV region, discussing
the several constraints that emerge from the model. In this case the
axion is, in general, not long-lived and as such is not a component of
dark matter. On the other hand, the constraints from CAST can be
avoided, since the particle would not be produced by ordinary thermal mechanisms at the center of the
sun, where the temperature is about 1.7 keV,  being its mass above the keV range. Obviously, in this case other
constraints emerge from nucleosynthesis requirements, since a particle
in this mass range has to decay fast enough in order not to generate a
late entropy release at nucleosynthesis time. We have seen that an
axion in the MeV range is consistent with these two requirements. An axion of this type could be searched for at colliders, and in this respect the analysis of its possible detection at
the LHC would follow quite closely the patterns described by two of us in
\cite{Coriano:2009zh}. As in this previous (non-supersymmetric) study, where the axion is Higgs-like (of a mass in the GeV region) typical channels where to look for this particle would be a) the associated
production of an axion and a direct photon; b) the multi-axion production channel,
and c) the associated production of one axion and other Higgses 
of the CP-even sector. The modifications, compared to that previous study, would now involve 1) the lower value of the mass of the axion (MeV rather than GeV); 2) the presence of extra supersymmetric interactions.

\subsection{Comments}
One important comment concerns the connection between these class of models and their completion theories such as string theory, which lay at their foundation. In our study we have selected a scenario characterized by low energy supersymmetry, with a phenomenological analysis that is essentially connected with the TeV scale and above. 
This is the scale which is likely to be scanned in the near future by several experiments, including the LHC, and for this reason we have directed out numerical studies in this direction. There is, however, a second 
 it involves a value of 
$M_{St}$ which is very large and close to the Planck scale. In this second case the model predicts, obviously, a decoupling of the anomalous symmetry, leaving at low energy a scenario which is essentially the same of the MSSM, since the extra $Z$ prime, which is part of the spectrum, is extremely heavy. This would obviously imply a decoupling both of the anomalous gauge boson and of the anomalous trilinear interactions which are associated with it. A physical axion could, however, survive this limit, if the scale of the extra potential is also of the order of $M_{St}$, but its interaction with ordinary matter would be extremely suppressed by the same scale. 

\section{Conclusions} 
The investigation of the phenomenological role
played by models containing anomalous gauge interactions from abelian
extensions of the Standard Model, we believe that will receive further
attention in the future. These studies can be motivated within several
scenarios, including string and supergravity theories, in which gauged
axionic symmetries are introduced for anomaly cancellation.  In turn,
these modified mechanisms of cancellation of the anomalies, which
involve an anomalous fermion spectrum and an axion, are essentially
connected with the UV completion of these field theories, which in a
string framework is realized by the Green-Schwarz mechanism.

The model that we have investigated (the USSM-A) summarizes the most
salient physical features of these types of constructions, where a
St\"uckelberg supermultiplet is associated to an anomalous abelian
structure in order to restore the gauge invariance of the anomalous
effective action. 
In this work we have tried to characterize in detail some among the
main phenomenological implications of these models, which are
particularly interesting for cosmology. The physical axion of this
construction, or gauged axion, emerges as a component of the
St\"uckelberg field $\textrm{Im}\, b$. We have pointed out that
the mechanism of sequential misalignment, formerly discussed in the
non-supersymmetric case \cite{Coriano:2010py}, finds a natural
application also in the presence of supersymmetry, with minor
modifications. 

One relevant feature of these models, already noticed in \cite{Coriano:2010py}, is that their axions do not contribute
to the isocurvature perturbations of the early universe, being gauge degrees of freedom at the scale of inflation. 

We have followed a specific pattern in order to come out with specific
results in these types of models, using for this purpose a particular
superpotential (the USSM superpotential), whose essential features, however, may well be generic.

We have presented an accurate study of the neutralino relic densities, showing that the St\"uckelberg mass value is constrained by the requirement of a consistent mass spectrum, with values for the lightest CP-even Higgs larger 
than the current LHC limits ( $>$ 120 GeV) 

and by the experimental value for the dark matter abundance from WMAP 
\cite{Jarosik:2010iu}.
Thus, in these models, the allowed value of the St\"uckelberg scale is positively correlated with the value of $\tan\beta$. As it grows, $\tan\beta$ has also to grow (for a fixed value of the vev of the singlet 
$v_S$) in order to preserve the WMAP bound. In particular $M_{St}$ and $v_S$ are positively correlated. This correlation is necessary in order to obtain values of the neutralino mass which allow to satisfy the same bounds, which in our case is around 20 GeV.

We have seen that with a St\"uckelberg mass in the TeV range
the non-thermal population of axions does not contribute
significantly to the dark matter densities if these axions are PQ-like. These types of
constraints, obviously, are typical of supersymmetric constructions
and are avoided in a non-supersymmetric context. In this second case,
as discussed in \cite{Coriano:2010py}, a St\"uckelberg scale around
$10^7$ GeV is sufficient to revert this trend. 

We have also pointed out that gauged axions in the milli-eV mass range 
are probably difficult to reconcile with current bounds from direct searches, while the case for detecting MeV or heavier axions, in these types of models, remains a wide open possibility. In this second case, cascade decays of these light particles and their associated production with photons should be seen as their possible event signatures at the LHC.

\centerline{\bf Acknowledgements} 
We thank Nikos Irges, George Lazarides and Antonio Racioppi for discussions. This work is supported in part by the European Union through the Marie Curie Research and Training Network UniverseNet (MRTN-CT-2006-035863).

\begin{appendix}

\section{General Features of the Model }
\label{thelagrangian}
In this appendix we summarize some of the basic features of the USSM-A. 
The gauge structure of the model is of the form
$SU(3)_c\times SU(2) \times U(1)_Y \times U(1)_B$, where $B$ is the
anomalous gauge boson, and with a matter content given by the usual
generations of the Standard Model (SM).  In all the Lagrangians below
we implicitly sum over the three fermion generations. A list of the
fundamental superfields and charge assignments is summarized in
Tab.~\ref{fieldcontent}. 
The Lagrangian can be expressed as
\begin{equation} {\cal L}_{USSM-A}={\cal L}_{USSM}+ {\cal L}_{KM}+
{\cal L}_{FI} + {\cal L}_{axion}
\end{equation} where the Lagrangian of the USSM (${\cal L}_{USSM}$)
has been modified by the addition of ${\cal L}_{axion}$ to compensate
for the anomalous variation of the corresponding effective action due
to the anomalous charge assignments. The former is given by
\begin{equation} {\cal L}_{USSM}={\cal L}_{lep}+{\cal L}_{quark}+{\cal
L}_{Higgs}+{\cal L}_{gauge}+ {\cal L}_{SMT}+{\cal L}_{GMT}
\end{equation} 
with contributions from the leptons, quarks and Higgs
plus gauge kinetic terms. The matter contributions from leptons and
quarks
\begin{align} 
&{\cal
L}_{lep}=\int{d^{4}\theta\left[\hat{L}^{\dagger}e^{2g_2\hat{W}
+g_Y\hat{Y}+g_B\hat{B}}\hat{L}+\hat{R}^{\dagger}e^{2g_2\hat{W}
+g_Y\hat{Y}+g_B\hat{B}}\hat{R}\right]}\\ &{\cal L}_{quark}=\int
d^{4}\theta\left[\hat{Q}^{\dagger}e^{2g_s\hat{G}+2g_2\hat{W}
+g_Y\hat{Y}+g_B\hat{B}}\hat{Q}+\hat{U}_{R}^{\dagger}e^{2g_s\hat{G}
+g_Y\hat{Y}+g_B\hat{B}}\hat{U}_{R}+\hat{D}_{R}^{\dagger}e^{2g_s\hat{G}
+g_Y\hat{Y}+g_B\hat{B}}\hat{D}_{R}\right]
\end{align} 
are accompanied by a sector which involves two Higgs
$SU(2)$ doublet superfields, $\hat{H}_{1}$ and $\hat{H}_{2}$, and one
singlet $\hat{S}$
\begin{align} {\cal L}_{Higgs} =
\int{d^{4}\theta\left[\hat{H}_{1}^{\dagger}e^{2g_2\hat{W}+g_Y\hat{Y}
+g_B\hat{B}}\hat{H}_{1}+\hat{H}_{2}^{\dagger}e^{2g_2\hat{W}+g_Y\hat{Y}
+g_B\hat{B}}\hat{H}_{2}+\hat{S}^{\dagger}e^{g_B\hat{B}}\hat{S} +{\cal
W}\delta^{2}(\bar{\theta})+\bar{{\cal W}}\delta^{2}(\theta)\right]}
\end{align} with the superpotential chosen of the form
\begin{align} {\cal W} = \lambda
\hat{S}\hat{H}_{1}\cdot\hat{H}_{2}+y_{e}\hat{H}_{1}
\cdot\hat{L}\hat{R}+y_{d}\hat{H}_{1}\cdot\hat{Q}\hat{D}_{R}
+y_{u}\hat{H}_{2}\cdot\hat{Q}\hat{U}_{R}.
\label{supi}
\end{align} 
This superpotential, as shown in
\cite{Coriano:2008xa,Coriano:2008aw}, allows a physical axion in the
spectrum. The gauge content plus the soft breaking terms in the form
of scalar mass terms (SMT) are identical to those of the USSM
\begin{align} 
&{\cal L}_{gauge}=\frac{1}{4}\int{d^{4}\theta\left[{\cal
G}^{\alpha}{\cal
G}_{\alpha}+W^{\alpha}W_{\alpha}+W^{Y\alpha}W^{Y}_{\alpha}
+W^{B\alpha}W^{B}_{\alpha}\right]\delta^{2}(\bar{\theta})+h.c.}
\nonumber\\ 
&{\cal L}_{SMT}=-\int
d^{4}\theta\,\delta^{4}(\theta,\bar{\theta})\,[M^{2}_{L}\hat{L}^{\dagger}\hat{L}
+m^{2}_{R}\hat{R}^{\dagger}\hat{R}+M^{2}_{Q}\hat{Q}^{\dagger}\hat{Q}
+m^{2}_{U}\hat{U}_{R}^{\dagger}\hat{U}_{R}+m^{2}_{D}\hat{D}_{R}^{\dagger}\hat{D}_{R}
\nonumber\\ &\hspace{2.5cm}+m_{1}^{2}\hat{H}_{1}^{\dagger}\hat{H}_{1}
+m_{2}^{2}\hat{H}_{2}^{\dagger}\hat{H}_{2}+m_{S}^{2}\hat{S}^{\dagger}\hat{S}
+(a_{\lambda}\hat{S}\hat{H}_{1}\cdot\hat{H}_{2}+h.c.)+(a_{e}\hat{H}_{1}\cdot\hat{L}\hat{R}+h.c.)
\nonumber\\
&\hspace{2.5cm}+(a_{d}\hat{H}_{1}\cdot\hat{Q}\hat{D}_{R}+h.c.)
+(a_{u}\hat{H}_{2}\cdot\hat{Q}\hat{U}_{R}+h.c.)].
\end{align} 
As usual, $M_{L},M_{Q},m_R,m_{U_R},m_{D_R},m_1,m_2,m_S$
are the mass parameters of the explicit supersymmetry breaking, while
$a_e,a_{\lambda},a_u,a_d$ are dimensionful coefficients.  The soft
breaking due to gaugino mass terms (GMT) now include a mixing mass
parameter $M_{YB}$
\begin{align} 
&{\cal L}_{GMT}=\int d^{4} \theta \left[ \frac{1}{2}
\left(M_{G}{\cal G}^{\alpha}{\cal G}_{\alpha} + M_{w}W^{\alpha
}W_{\alpha} + M_YW^{Y\alpha} W^{Y}_{\alpha} + M_B W^{B\alpha}
W^{B}_{\alpha} \right. \right. \nonumber \\ &\qquad\qquad\qquad\qquad
\left. \left. + M_{Y B} W^{Y\alpha} W^{B}_{\alpha} \right)
+h.c.\right] \delta^{4}(\theta,\bar{\theta}).
\end{align} 
\\
\begin{table}
\begin{center}
\begin{tabular}{|l||c|c|c|c|} \hline Superfields &SU(3)& SU(2)&
$U(1)_{Y}$ & $U(1)_{B}$\\ \hline $\hat{\bf b}(x,\theta,\bar{\theta})$&
{\bf 1} & {\bf 1} & 0 & $ s $\\ $\hat{S}(x,\theta,\bar{\theta})$& {\bf
1} & {\bf 1} & 0 & $B_{S}$\\ $\hat{L}(x,\theta,\bar{\theta})$& {\bf 1}
& {\bf 2} & -1/2 & $B_{L}$\\ $\hat{R}(x,\theta,\bar{\theta})$& {\bf 1}
& {\bf 1} & 1 & $B_{R}$\\ $\hat{Q}(x,\theta,\bar{\theta})$& {\bf 3} &
{\bf 2} & 1/6 & $B_{Q}$\\ $\hat{U}_{R}(x,\theta,\bar{\theta})$&
$\bar{{\bf 3}}$ & {\bf 1} & -2/3 & $B_{U_R}$\\
$\hat{D}_{R}(x,\theta,\bar{\theta})$& $\bar{{\bf 3}}$ & {\bf 1} & +1/3
& $B_{D_R}$\\ $\hat{H}_{1}(x,\theta,\bar{\theta})$& {\bf 1} & {\bf 2}
& -1/2 & $B_{H_{1}}$\\ $\hat{H}_{2}(x,\theta,\bar{\theta})$& {\bf 1} &
{\bf 2} & 1/2 & $B_{H_{2}}$\\ \hline
\end{tabular}
\label{cariche}
\end{center}
\caption{Charge assignment of the model}
\label{fieldcontent}
\end{table} The superfield $\hat{\bf b}$ describes the St\"uckelberg
multiplet,
\begin{align} 
\hat{{\bf b}}= b + \sqrt{2}\theta\psi_{{\bf b}}
- i\theta\sigma^{\mu}\bar{\theta}\partial_{\mu} b
+ \frac{i}{\sqrt{2}}\theta\theta\bar{\theta}\bar{\sigma}^{\mu}\partial_{\mu}\psi_{{\bf
b}} -\frac{1}{4}\theta\theta\bar{\theta}\bar{\theta}\Box b 
- \theta\theta F_{{\bf b}},
\end{align} 
and contains the St\"uckelberg axion (a complex $b$ field)
and its supersymmetric partner, referred to as the axino ($\psi_{\bf
b}$), which combines with the neutral gauginos and higgsinos to
generate the neutralinos of the model. Details on the notation for the
superfields components can be found in Tab.~\ref{superfieldcomp}.  We just recall that 
we denote with $\lambda_B$ and $\lambda_Y$ the two gauginos of the two vector superfields $(\hat{B}, \hat{Y})$ 
corresponding to the anomalous $U(1)_B$ and to the hypercharge vector multiplet.  
The singlet superfield $\hat{S}$ has as components the scalar ``singlet'' $S$ and its supersymmetric partner, the singlino, 
denoted as $\tilde{S}$. 

The interactions and dynamics of the axion superfield are defined in
$\mathcal{L}_{axion}$, the Lagrangian that contains both the kinetic
(St\"uckelberg) term, responsible for the mass of the anomalous gauge
boson (which reaches the electroweak symmetry breaking scale already
in a massive state), the kinetic term of the saxion and of the axino,
and the Wess-Zumino terms, which are needed for anomaly
cancellation. We recall that St\"uckelberg fields appear both in anomalous and in non-anomalous contexts. The second one has been analyzed recently in \cite{Feldman:2010wy}.
\begin{table}
\begin{center}
\begin{tabular}{|l||c|c|c|} \hline Superfield & Bosonic & Fermionic &
Auxiliary \\ \hline $\hat{\bf b}(x,\theta,\bar{\theta})$& $b(x)$ &
$\psi_{\bf b}(x)$ & $F_{\bf b}(x)$ \\ $\hat{S}(x,\theta,\bar{\theta})$
& $S(x)$ & $\tilde{S}(x)$ & $F_{S}(x)$ \\
$\hat{L}(x,\theta,\bar{\theta})$ & $\tilde{L}(x)$ & $L(x)$ &
$F_{L}(x)$\\ $\hat{R}(x,\theta,\bar{\theta})$ & $\tilde{R}(x)$ &
$\bar{R}(x)$ & $F_{R}(x)$ \\ $\hat{Q}(x,\theta,\bar{\theta})$ &
$\tilde{Q}(x)$ & $Q(x)$ & $F_{Q}(x)$ \\
$\hat{U}_{R}(x,\theta,\bar{\theta})$& $\tilde{U}_R(x)$ &
$\bar{U}_R(x)$ & $F_{U_R}(x)$ \\ $\hat{D}_{R}(x,\theta,\bar{\theta})$&
$\tilde{D}_R(x)$ & $\bar{D}_R(x)$ & $F_{D_R}(x)$ \\
$\hat{H}_{1}(x,\theta,\bar{\theta})$& $H_1(x)$ & $\tilde{H_1}(x)$ &
$F_{H_1}(x)$ \\ $\hat{H}_{2}(x,\theta,\bar{\theta})$& $H_2(x)$ &
$\tilde{H_2}(x)$ & $F_{H_2}(x)$ \\ $\hat{B}(x,\theta,\bar{\theta})$ &
$B_{\mu}(x)$ & $\lambda_{B}(x)$ & $D_B(x)$ \\
$\hat{Y}(x,\theta,\bar{\theta})$ & $A^{Y}_{\mu}(x)$ & $\lambda_{Y}(x)$
& $D_Y(x)$ \\ $\hat{W}^{i}(x,\theta,\bar{\theta})$& $W^{i}_{\mu}(x)$ &
$\lambda_{W^{i}}(x)$ & $D_{W^{i}}(x)$ \\
$\hat{G}^{a}(x,\theta,\bar{\theta})$& $G^{a}_{\mu}(x)$ &
$\lambda_{g^a}(x),\bar{\lambda}_{g^a}(x)$ & $D_{G^{a}}(x)$ \\ \hline
\end{tabular}
\end{center}
\caption{Superfields and their components.}
\label{superfieldcomp}
\end{table}

The extra contributions to ${\cal L}_{axion}$, called ${\cal L}_{axion,i}$ are given by 
\begin{align}
 {\cal L}_{axion,i}= &M_{St}\textrm{Re}b D_B + M_{st} (i \psi_{\bf b}\lambda_{B}+h.c.)+ \displaybreak[0]\notag\\ 
&\frac{c_G}{M_{St}}\Big(
\frac{1}{16}F_{\bf b}\lambda_G^a \lambda_G^a+
\frac{i}{8\sqrt{2}} D_G^a \lambda_G^a\psi_{\bf b}-
\frac{1}{16\sqrt{2}}G^a_{\mu\nu}\psi_{\bf b}\sigma^\mu\bar{\sigma}^\nu\lambda_G^a+
\frac{1}{64\sqrt{2}}\textrm{Im}b\;\tilde{G}^{a\;\mu\nu}G^a_{\mu\nu}+
\notag\\
&
\frac{1}{8\sqrt{2}}\textrm{Im}b\;\lambda^a_G\sigma_\mu D^\mu\bar{\lambda}^a_G+
\frac{1}{16\sqrt{2}}\textrm{Re}b\;D_G^a D_G^a-\frac{1}{16\sqrt{2}}\textrm{Re}b\;G^{a\;\mu\nu}G^a_{\mu\nu}-
\frac{i}{8\sqrt{2}}\textrm{Re}b\;\lambda^a_G\sigma_\mu D^\mu\bar{\lambda}^a_G
+h.c.
\Big)+\displaybreak[0]\notag\\
&\frac{c_W}{M_{St}}\Big(
\frac{1}{16}F_{\bf b}\lambda_W^i \lambda_W^i+
\frac{i}{8\sqrt{2}} D_W^i \lambda_W^i\psi_{\bf b}-
\frac{1}{16\sqrt{2}}W^i_{\mu\nu}\psi_{\bf b}\sigma^\mu\bar{\sigma}^\nu\lambda_W^i+
\frac{1}{64\sqrt{2}}\textrm{Im}b\;\tilde{W}^{i\;\mu\nu}W^i_{\mu\nu}+
\notag\\
&
\frac{1}{8\sqrt{2}}\textrm{Im}b\;\lambda^i_W\sigma_\mu D^\mu\bar{\lambda}^i_W+
\frac{1}{16\sqrt{2}}\textrm{Re}b\;D_W^i D_W^i-
\frac{1}{16\sqrt{2}}\textrm{Re}b\;W^{i\;\mu\nu}W^i_{\mu\nu}-
\frac{i}{8\sqrt{2}}\textrm{Re}b\;\lambda^i_W\sigma_\mu D^\mu\bar{\lambda}^i_W
+h.c.
\Big)+\displaybreak[0]\notag\\
&\frac{c_Y}{M_{St}}\Big(
\frac{1}{2}F_{\bf b}\lambda_Y \lambda_Y+
\frac{i}{\sqrt{2}} D_Y \lambda_Y \psi_{\bf b}-
\frac{1}{2\sqrt{2}}F^Y_{\mu\nu}\psi_{\bf b}\sigma^\mu\bar{\sigma}^\nu\lambda_Y+
\frac{1}{8\sqrt{2}}\textrm{Im}b\;\tilde{F}^{Y\;\mu\nu}F^Y_{\mu\nu}-
\frac{1}{\sqrt{2}}\textrm{Im}b\;\lambda_Y\sigma_\mu \partial^\mu\bar{\lambda}_Y +
\notag\\
&\frac{1}{2\sqrt{2}}\textrm{Re}b\;D_Y D_Y -
\frac{1}{2\sqrt{2}}\textrm{Re}b\;F^{Y\;\mu\nu}F^Y_{\mu\nu}-
\frac{i}{\sqrt{2}}\textrm{Re}b\;\lambda_Y\sigma_\mu \partial^\mu\bar{\lambda}_Y
+h.c.
\Big)+\displaybreak[0]\notag\\
&\frac{c_B}{M_{St}}\Big(
\frac{1}{2}F_{\bf b}\lambda_B \lambda_B+
\frac{i}{\sqrt{2}} D_B \lambda_B \psi_{\bf b}-
\frac{1}{2\sqrt{2}}F^B_{\mu\nu}\psi_{\bf b}\sigma^\mu\bar{\sigma}^\nu\lambda_B+
\frac{1}{8\sqrt{2}}\textrm{Im}b\;\tilde{F}^{B\;\mu\nu}F^B_{\mu\nu}-
\frac{1}{\sqrt{2}}\textrm{Im}b\;\lambda_B\sigma_\mu \partial^\mu\bar{\lambda}_B +
\notag\\
&\frac{1}{2\sqrt{2}}\textrm{Re}b\;D_B D_B -
\frac{1}{2\sqrt{2}}\textrm{Re}b\;F^{B\;\mu\nu}F^B_{\mu\nu}-
\frac{i}{\sqrt{2}}\textrm{Re}b\;\lambda_B\sigma_\mu \partial^\mu\bar{\lambda}_B
+h.c.
\Big)+\displaybreak[0]\notag\\
&\frac{c_{YB}}{M_{St}}\Big(
-\frac{1}{2}F_{\bf b}\lambda_B \lambda_Y-
\frac{i}{2\sqrt{2}} D_Y \lambda_B \psi_{\bf b}-
\frac{i}{2\sqrt{2}} D_B \lambda_Y \psi_{\bf b}+
\frac{1}{4\sqrt{2}}F^B_{\mu\nu}\psi_{\bf b}\sigma^\mu\bar{\sigma}^\nu\lambda_Y+
\frac{1}{4\sqrt{2}}F^Y_{\mu\nu}\psi_{\bf b}\sigma^\mu\bar{\sigma}^\nu\lambda_B-
\notag\\
&\frac{1}{8\sqrt{2}}\textrm{Im}b\;\tilde{F}^{Y\;\mu\nu}F^B_{\mu\nu}+
\frac{1}{2\sqrt{2}}\textrm{Im}b\;\lambda_Y\sigma_\mu \partial^\mu\bar{\lambda}_B+
\frac{1}{2\sqrt{2}}\textrm{Im}b\;\lambda_B\sigma_\mu \partial^\mu\bar{\lambda}_Y -
\notag\\
&\frac{1}{2\sqrt{2}}\textrm{Re}b\;D_Y D_B +
\frac{1}{2\sqrt{2}}\textrm{Re}b\;F^{Y\;\mu\nu}F^B_{\mu\nu}-
\frac{i}{2\sqrt{2}}\textrm{Re}b\;\lambda_Y\sigma_\mu \partial^\mu\bar{\lambda}_B-
\frac{i}{2\sqrt{2}}\textrm{Re}b\;\lambda_B\sigma_\mu \partial^\mu\bar{\lambda}_Y
+h.c.
\Big).
\label{laxion}
\end{align}

We finally recall that the three scalar sectors of the model are
characterized in terms of
\begin{itemize}
\item{A {\bf Charged Higgs sector}}\\ 
This sector involves the states
$(\textrm{Re}H_2^{1},\textrm{Re}H_1^{2})$. The mass matrix has one zero eigenvalue corresponding to a charged Goldstone boson and a mass eigenvalue corresponding to the charged Higgs mass
\begin{align}
m^{2}_{H^\pm}=\left(\frac{v_{1}}{v_{2}}+\frac{v_{2}}{v_{1}}\right)
\left(\frac{1}{4}g^{2}v_{1}v_{2}-\frac{1}{2}\lambda^{2}v_{1}v_{2}
+a_{\lambda}\frac{v_{S}}{\sqrt{2}}\right);
\end{align} 
where $g^2=g_2^2+g_Y^2$.

\item{A {\bf CP-even sector}}\\ 
This sector is diagonalized starting
from the basis $(\textrm{Re}H_1^1,\textrm{Re}H_2^2,\textrm{Re}S,\textrm{Re}b)$
The four physical states obtained in this sector are
denoted, as $H_0^1, H_0^2, H_0^3$ and $H_0^4$. Together with the charged
physical state extracted before, $H^{\pm}$, they describe the 6
degrees of freedom of the CP-even sector.

\item{A {\bf CP-odd sector}}\\ 
This sector is diagonalized starting
from the basis $(\textrm{Im}H_1^1,\textrm{Im}H_2^2,\textrm{Im}S,
\textrm{Im}b)$.  We obtain two physical states, $H_0^4$ and $H_0^5$,
and two Goldstone modes that provide the longitudinal degrees of
freedom for the neutral gauge bosons, $Z$ and $Z^{\prime}$.
\end{itemize}

\section{Neutralino mass matrix}
\label{sec:neutralino} 
Now we turn to the neutralino sector; the mass matrix in the basis
$(i\lambda_{w_3},i\lambda_{Y},i\lambda_{B},\tilde{H}_1^1,\tilde{H}_2^2,\tilde{S},\psi_{\bf
b})$ takes the form
\begin{equation} M_{\chi^0}=
\begin{pmatrix} M_{\chi^0}^{\,\,11} & 0 & 0 & M_{\chi^0}^{\,\,14} &
M_{\chi^0}^{\,\,15} & 0 & M_{\chi^0}^{\,\,17}\\ \cdot &
M_{\chi^0}^{\,\,22} & M_{\chi^0}^{\,\,23} & M_{\chi^0}^{\,\,24} &
M_{\chi^0}^{\,\,25} & 0 & 0\\ \cdot & \cdot &
M_{\chi^0}^{\,\,33} & M_{\chi^0}^{\,\,34} & M_{\chi^0}^{\,\,35} &
M_{\chi^0}^{\,\,36} & M_{\chi^0}^{\,\,37}\\ \cdot & \cdot & \cdot & 0
& M_{\chi^0}^{\,\,45} & M_{\chi^0}^{\,\,46} & 0\\ \cdot & \cdot &
\cdot & \cdot & 0 & M_{\chi^0}^{\,\,56} & 0\\ \cdot & \cdot & \cdot &
\cdot & \cdot & 0 & 0\\ \cdot & \cdot & \cdot & \cdot & \cdot & \cdot
& M_{\chi^0}^{\,\,77}\\
\end{pmatrix}
\end{equation} 
with
\begin{align} 
M_{\chi^0}^{\,\,11} &= M_{w} \hspace{1cm}
M_{\chi^0}^{\,\,14} = -\frac{g_2 v_1}{2} \hspace{1cm}
M_{\chi^0}^{\,\,15} = \frac{g_2 v_2}{2} \hspace{1cm}
M_{\chi^0}^{\,\,17} = 0
\nonumber\displaybreak[0]\\ 
M_{\chi^0}^{\,\,22} &=M_{Y} 
\hspace{1cm} 
M_{\chi^0}^{\,\,23} = \frac{1}{2} M_{YB}
\hspace{1cm} 
M_{\chi^0}^{\,\,24} = \frac{g_Y v_1}{2}
\hspace{1cm} 
M_{\chi^0}^{\,\,25} = -\frac{g_Y v_2}{4} 
\nonumber\displaybreak[0]\\ 
M_{\chi^0}^{\,\,33} &=\frac{1}{2} M_B
\hspace{1cm} 
M_{\chi^0}^{\,\,34} = -v_1 g_B B_{H_1}  \hspace{1cm} 
M_{\chi^0}^{\,\,35} = -v_2 g_B B_{H_2} \hspace{1cm} 
M_{\chi^0}^{\,\,36} = -v_S g_B B_S  \nonumber\displaybreak[0]\\ 
M_{\chi^0}^{\,\,37} &= M_{St} \hspace{1cm} 
M_{\chi^0}^{\,\,45} = \frac{\lambda v_S}{\sqrt{2}} \hspace{1cm} 
M_{\chi^0}^{\,\,46} = \frac{\lambda v_2}{\sqrt{2}} \hspace{1cm} 
M_{\chi^0}^{\,\,56} = \frac{\lambda v_1}{\sqrt{2}} \hspace{1cm} 
M_{\chi^0}^{\,\,77} = - M_{b}
\end{align} The rotation matrix for this sector is implicitly defined
as $O^{\chi^0}$ and
\begin{equation}
\begin{pmatrix} 
i\lambda_{w_3}\\ 
i\lambda_{Y}\\ 
i\lambda_{B}\\
\tilde{H}_1^1\\ 
\tilde{H}_2^2 \\
\tilde{S}\\ 
\psi_{\bf b}
\end{pmatrix}= 
O^{\chi^0}
\begin{pmatrix} 
\chi^0_0\\ 
\chi^0_1\\ 
\chi^0_2\\ 
\chi^0_3 \\ 
\chi^0_4\\ 
\chi^0_5 \\ 
\chi^0_6
\end{pmatrix}.
\end{equation}

\subsection{Chargino sector}
\label{sec:chargino} We recall here the structure of the chargino
sector and the diagonalization procedure.  We define
\begin{equation} \lambda_{w^+}=\frac{1}{\sqrt{2}}(\lambda_{w_1} - i
\lambda_{w_2})\hspace{1cm}\lambda_{w^-}=\frac{1}{\sqrt{2}}(\lambda_{w_1}
+ i \lambda_{w_2})
\end{equation} and in the basis
$\left(\lambda_{w^+},\tilde{H}_{2}^{1},\lambda_{w^-},\tilde{H}_{1}^{2}\right)$
we obtain the mass matrix
\begin{equation} M_{\tilde{\chi}^{\pm}}^{2}=
\begin{pmatrix} 0 & 0 & M_W & g_2 v_1\\ 0 & 0 & g_2 v_2 & \lambda
v_S\\ M_W & g_2 v_2 & 0 & 0\\ g_2 v_1 & \lambda v_S & 0 & 0
\end{pmatrix};
\end{equation} from the diagonalization we get the squared eigenvalues
\begin{equation} m_{\tilde{\chi}^{\pm}_{1,2}}=\frac{1}{2}\left[M_W^{2}
+ \lambda^{2}v_S^2 + g_2^2 v^2 \mp \sqrt{\left(M_W^{2} +
\lambda^{2}v_S^2 + g_2^2 v^2\right)^{2} - 4\left(\lambda v_S M_W
-g_2^2 v_1 v_2\right)^2}\right]
\end{equation} If we define
\begin{equation} \psi^{+}= \left(
\begin{array}{c} \lambda_{w^+}\\ \tilde{H}_{2}^{1}
\end{array} \right)\hspace{1cm} \psi^{-}= \left(
\begin{array}{c} \lambda_{w^-}\\ \tilde{H}_{1}^{2}
\end{array} \right)
\end{equation} and define the mass eigenstates as
\begin{equation} \chi^{+}=V\psi^{+}\hspace{1cm}\chi^{-}=U\psi^{-}
\end{equation} where $U$ and $V$ are two unitary matrices that perform
the diagonalization of this sector. If we define
\begin{equation} X=
\begin{pmatrix} M_W & g_2 v_2\\ g_2 v_1 & \lambda v_S
\end{pmatrix}
\end{equation} then these unitary matrices are defined is such a way
that
\begin{equation} V X^\dagger X V^{-1}=U^* X X^\dagger U^T =
M_{\chi^{\pm}, diag};
\end{equation} where $M_{\chi^{\pm}, diag}$ is given by
\begin{equation} M_{\chi^{\pm}, diag}= \left(
\begin{array}{cc} m_{\tilde{\chi}^{\pm}_{1}} & 0\\ 0 &
m_{\tilde{\chi}^{\pm}_{2}}
\end{array} \right).
\end{equation}

\section{Appendix. Relic densities at the second misalignment}
\label{relics} 
In this appendix we fill in the gaps in the derivation
of the expression of the abundances generated by the mechanism of
vacuum misalignment. We start from the Lagrangian 
\begin{align} 
\mathcal{S}=
\int d^4 x \sqrt{g}\left( \frac{1}{2}\dot{\chi}^2 - \frac{1}{2}
m_\chi^2 \Gamma_\chi \dot{\chi}\right), 
\end{align} 
where $\Gamma_\chi$ is
the decay rate of the axion and we have expanded the potential around
its minimum up to quadratic terms. The same action is derived from the
quadratic approximation to the general expression 
\begin{align}
\mathcal{S}=\int d^4 x R^3(t)\left( \frac{1}{2} \sigma_{\chi}^2
\left(\partial_{\alpha} \theta\right)^2 - \mu^4 \left(1-
\cos\theta\right) - V_0\right) 
\end{align} 
which, in our case, is constructed
from the expression of $V^\prime$ given in Eq.~(\ref{extrap}), with
$\mu\sim v$, the electroweak scale. We also set other contributions to
the vacuum potential to vanish ($V_0=0$). In a
Friedmann-Robertson-Walker spacetime metric with a scaling factor
$R(t)$, this action gives the equation of motion 
\begin{align}
\frac{d}{dt}\biggl[\left( R^3(t) (\dot{\chi} +
\Gamma_\chi\right)\biggr] + R^3 m_{\chi}^2(T) =0.
\label{FRWequation}  
\end{align}
We will neglect the decay rate of the axion
in this case and set $\Gamma_\chi\approx 0$. At this point, since the
potential $V^\prime$ is of non-perturbative origin, we can assume that
it vanishes far above the electroweak scale (or temperature
$T_{ew}$). For this reason $m_\chi=m_b=0$ for $T\gg T_{ew}$, which is
essentially equivalent to assume that the St\"uckelberg axion is not
subject to any mixing far above the weak scale.  The general equation
of motion derived from Eq.~(\ref{FRWequation}), introducing a
temperature dependent mass, can be written as
\begin{align} 
\ddot{\chi} + 3 H \dot{\chi} + m_{\chi}^2(T) \chi =0,
\label{motioneq}
\end{align} 
which clearly allows as a solution a constant value of
the misalignment angle $\theta=\theta_{i}$. The T-dependence of the
mass term should be generated, for consistency, from a generalization
to finite temperature of $V^\prime$. In practice this is not necessary
in our case, being the role of the first misalignment negligible in
determining the final mass of the axion.\\
The axion energy density is given by 
\begin{align}
\rho=\frac{1}{2}
\dot{\chi}^2 + \frac{1}{2} m_\chi^2 \chi^2,
\label{rhoeq}  
\end{align}
which after a harmonic averaging gives 
\begin{align}
\langle
\rho\rangle = m_\chi^2 \langle \chi^2\rangle.
\label{averageeq}  
\end{align}
Notice that after differentiating Eq.~(\ref{rhoeq}) and using 
the equation of motion in~(\ref{motioneq}), followed by the 
averaging Eq.~(\ref{averageeq}) one obtains the relation
\begin{align} 
\langle \dot{\rho}\rangle =\langle \rho\rangle \left(
-3 H + \frac{\dot{m}}{m}\right),
\end{align} 
where the time dependence of the mass is through its
temperature $T(t)$, while $H(t)=\dot{R}(t)/R(t)$ is the Hubble
parameter. By inspection one easily finds that the solution of this
equation is of the form
\begin{align} 
\langle \rho\rangle = \frac{m_\chi(T)}{R^3(t)}
\end{align} 
showing a dilution of the energy density with an
increasing space volume, valid even for a $T$-dependent mass. At this
point, the universe must be (at least) as old as the required period
of oscillation in order for the axion field to start oscillating and
to appear as dark matter, otherwise $\theta$ is misaligned but frozen;
this is the content of the condition
\begin{align} 
m_\chi(T_i)= 3 H(T_i),
\label{mhcond}
\end{align} which allows to identify the initial temperature of the
coherent oscillation of the axion field $\chi$, $T_i$, by equating
$m_\chi(T)$ to the Hubble rate, taken as a function of temperature.

To quantify the relic densities at the current temperature $T_0$
($T_0\equiv T(t_0)$, at current time $t_0$) we define preliminarily
the two standard effective couplings
\begin{align} 
& g_{*,S,T}=\sum_B g_i \left(\frac{T_i}{T}\right)^3 +
\frac{7}{8}\sum_F g_i \left(\frac{T_i}{T}\right)^3 \nonumber \\ 
& g_{*,T}=\sum_B g_i \left(\frac{T_i}{T}\right)^4 + \frac{7}{8}\sum_F g_i \left(\frac{T_i}{T}\right)^4,
\end{align} 
functions of the massless relativistic degrees of freedom
of the primordial state, with $T\gg T_{ew}$. The counting of the
degrees of freedom is: 2 for a Majorana fermion and for a massless
gauge boson, 3 for a massive gauge boson and 1 for a real scalar. In
the radiation era, the thermodynamics of all the components of the
primordial state is entirely determined by the temperature $T$, being
the system at equilibrium. We exclude for simplicity all sorts of
possible source of entropy due to any inhomogeneity (see, for
instance, \cite{Lazarides:1990xp}). Pressure and entropy are then just
given as a function of the temperature
\begin{align} & \rho=3 p=\frac{\pi^2}{30} g_{*,T}T^4 \nonumber \\ &
s=\frac{2 \pi^2}{45} g_{*,S,T} T^3,
\label{entropy}
\end{align} while the Friedmann equation allows to relate the Hubble
parameter and the energy density
\begin{equation} H=\sqrt{\frac{8}{3} \pi G_N \rho},
\label{hubble}
\end{equation} with $G_N={1}/{M_P^2}$ being the Newton constant and
$M_P$ the Planck mass. The number density of axions $n_\chi$ decreases
as $1/R^3$ with the expansion, as does the entropy density $s\equiv
S/R^3$, where $S$ indicates the comoving entropy density - which
remains constant in time ($\dot{S}=0$) - leaving the ratio $Y_a\equiv
n_\chi/s$ conserved. We define, as usual, the abundance variable of
$\chi$
\begin{equation} Y_\chi(T_i)= \frac{n_\chi}{s}\bigg\vert_{T_i}
\end{equation} at the temperature of oscillation $T_i$, and observe
that at the beginning of the oscillations the total energy density is
just the potential one
\begin{equation} \rho_\chi=n_\chi(T_i) m_\chi(T_i)=1/2
m_\chi^2(T_i)\chi_i^2.
\end{equation} We then obtain for the initial abundance at $T=T_i$
\begin{equation} Y_\chi(T_i)= \frac{1}{2}\frac{m_\chi(T_i)
\chi_i^2}{s}= \frac{45 m_\chi(T_i)\chi_i^2}{4 \pi^2 g_{*,S,T} T_i^3}
\label{yeq}
\end{equation} where we have inserted at the last stage the expression
of the entropy of the system at the temperature $T_i$ given by
Eq.~(\ref{entropy}).  At this point, plugging the expression of $\rho$
given in Eq.~(\ref{entropy}) into the expression of the Hubble rate as
a function of density given n Eq.~(\ref{hubble}), the condition for
oscillation Eq.~(\ref{mhcond}) allows to express the axion mass at
$T=T_i$ in terms of the effective massless degrees of freedom
evaluated at the same temperature, that is
\begin{equation} m_\chi(T_i)=\sqrt{\frac{4}{5}\pi^3
g_{*,T_i}}\frac{T_i^2}{M_P}.
\label{Tmass}
\end{equation} This gives for Eq.~(\ref{yeq}) the expression
\begin{equation} Y_\chi(T_i)= \frac{45
\sigma_\chi^2\theta_i^2}{2\sqrt{5 \pi g_{*, T_i}} T_i M_P},
\label{ychi}
\end{equation} where we have expressed $\chi$ in terms of the angle of
misalignment $\theta_i$ at the temperature when oscillations start. We
assume that $\theta_i=\langle \theta\rangle$ is the zero mode of the
initial misalignment angle after an averaging.  As we have already
mentioned, $T_i$ should be determined consistently by
Eq.~(\ref{mhcond}). However, the presence of two significant and
unknown variables in the expression of $m_\chi$, which are the
coupling of the anomalous $U(1)$, $g_B$, and the St\"uckelberg mass
$M$, forces us to consider the analysis of the T-dependence of $\chi$
phenomenologically less relevant. It is more so if the St\"uckelberg
mass is somehow close to the TeV region, in which case the zero
temperature axion mass $m_\chi$ acquires corrections proportional to
the bare coupling $\left( m_\chi\sim \lambda v(1 + O(g_B)\right)$.

For this reason, assuming that the oscillation temperature $T_i$ is
close to the electroweak temperature $T_{ew}$, Eq.~(\ref{Tmass})
provides an upper bound for the mass of the axion at which the
oscillations occur, assuming that they start around the electroweak
phase transition. Stated differently, mass values of $\chi$ such that
$m(T_i)\ll 3 H(T_i)$ correspond to frozen degrees of freedom of the
axion at the electroweak scale. This is clearly an approximation, but
it allows to define the oscillation mass in terms of the Hubble
parameter for each given temperature.

We recall that the relic density due to misalignment can be extracted
from the relations
\begin{equation} \Omega_\chi^{mis}\equiv \frac{\rho_{\chi
0}^{mis}}{\rho_c} =\frac{(n_{\chi 0}
m_\chi)}{\rho_c}=\left(\frac{n_{\chi\,0}}{s_0}\right)\frac{m_\chi
s_0}{\rho_c}
\label{resid}
\end{equation} where we have denoted with $n_{\chi 0}$ the current
number density of axions and with $\rho^{mis}_{\chi 0}$ their current
energy density due to vacuum misalignment. This expression can be
rewritten as
\begin{equation} \Omega_\chi^{mis}=\frac{n_\chi}{s}\bigg\vert_{T_i}
m_\chi\frac{s_0}{\rho_c}
\label{omegaeq}
\end{equation} using the conservation of the abundance $Y_{a
0}=Y_{a}(T_i)$. Notice that in Eq.~(\ref{omegaeq}) we have neglected a
possible dilution factor $\gamma=s_{osc}/s_0$ which may be present due
to entropy release. We have introduced the variable
\begin{equation} \rho_c= \frac{3 H_0^2}{8 \pi G_N}
\end{equation} which is the critical density and
\begin{equation} s_0= \frac{2 \pi^2}{45} g_{* S, T_0} T_0^3
\end{equation} which is the current entropy density. To fix $g_{* S,
T_0}$ we just recall that at the current temperature $T_0$ the
relativistic species contributing to the entropy density $s_0$ are the
photons and three families of neutrinos with
\begin{equation} g_{* S, T_0}= 2 + \frac{7}{8} \times 3\times 2
\left(\frac{T_\nu}{T_0}\right)^3
\end{equation} where, from entropy considerations,
${T_\nu}/{T_0}=(4/11)^{1/3}$.

To proceed with the computation of the massless degrees of freedom
above the electroweak phase transition we just recall the structure of
the model. We have 13 gauge bosons corresponding to the gauge group
$SU(3)\times SU(2)\times U_Y(1)\times U_B(1)$, 2 Higgs doublets, 3
generations of leptons and 3 families of quarks. Above the energy of
the electroweak transition we have only massless fields with the
exception of the $U_B(1)$ gauge boson, since this symmetry takes the
St\"uckelberg form above the electroweak scale, giving
$g_{*,T}=110.75$. Below the same scale this number is similarly
computed with $g_{*,T}=91.25$. Other useful parameters are the
critical density and the current entropy
\begin{equation}
\rho_{c}=5.2\cdot10^{-6}\textrm{GeV}/\textrm{cm}^3\hspace{1cm}s_0=2970
\,\,\textrm{cm}^{-3},
\end{equation} with $\theta_i\simeq1$.  It is clear, by inserting
these numbers into Eq.~(\ref{resid}) that $\Omega_\chi^{mis}$ \beq
\frac{45 \sigma_\chi^2\theta_i^2}{2\sqrt{5 \pi g_{*, T_i}} T_i M_P}
\frac{m_\chi s_0}{\rho_c} \eeq is negligible unless $\sigma_\chi\sim
M_{St}^2/v$ is of the same order of $f_a\sim 10^{12}$ GeV, the
standard PQ constant.  This choice would correspond to
$\Omega_\chi\sim 0.1$, but the value of $M_{St}$ should be of
$O(10^7)$ GeV
.

\end{appendix}

\end{document}